\pdfoutput=1
\documentclass{aa}
\usepackage{grffile}
\usepackage{graphicx}
\usepackage{txfonts}
\usepackage{mathtools}
\usepackage{lscape}

\bibpunct{(}{)}{;}{a}{}{,}

\newcommand{\dif}{$\rm cm^{2}~s^{-1}$}

\def\llm{\textsc{LLmodels}}
\def\phoenix{\textsc{PHOENIX}}

\def\synth3{\textsc{Synth3}}

\def\taurex{$\tau$-REx}
\def\helios{\textsc{HELIOS}}
\def\heliosk{\textsc{HELIOS-K}}
\def\vulcan{\textsc{VULCAN}}
\def\fastchem{\textsc{FastChem}}
\def\atmo{\textsc{ATMO}}

\def\Rsun{R_{\odot}}

\def\teff{T_{\rm eff}}
\def\teq{T_{\rm eq}}

\def\lalpha{\mathrm{Ly-}\alpha}

\def\mum{$\mu$m}

\def\hho{H$_{\rm 2}$O}
\def\co{CO}

\def\coo{CO$_{\rm 2}$}

\def\chhhh{CH$_{\rm 4}$}
\def\nhhh{NH$_{\rm 3}$}
\def\hcn{HCN}
\def\hh{H$_{\rm 2}$}
\def\nn{N$_{\rm 2}$}
\def\hminus{H$^-$}
\def\heminus{He$^-$}
\def\electron{$e^-$}

\begin{document}

\title{Stellar impact on disequilibrium chemistry and on observed spectra of hot Jupiter atmospheres}

\author{
D. Shulyak\inst{1}
\and
L.~M. Lara\inst{2}
\and
M. Rengel\inst{1}
\and
N.-E. N\`emec\inst{1}
}

\institute{
Max-Planck Institut f\"ur Sonnensystemforschung, Justus-von-Liebig-Weg 3, D-37077, G\"ottingen, Germany\\
\email{shulyak@mps.mpg.de}
\and
Instituto de Astrof\'{\i}sica de Andaluc\'{\i}a - CSIC, c/ Glorieta de la Astronom\'{\i}a s/n, 18008 Granada, Spain.}

\date{Received ; accepted}

\abstract
{}
{In this work we study the effect of disequilibrium processes (photochemistry and vertical transport) on mixing ratio 
profiles of neutral species and on the simulated
spectra of a hot Jupiter exoplanet that orbits stars of different spectral types. We additionally
address the impact of stellar activity that should be present to a different degree in all stars with convective envelopes.}
{We used the \vulcan\ chemical kinetic code to compute number densities of species in irradiated planetary atmospheres. 
The temperature-pressure profile of the atmosphere was computed with the \helios\ code.
We also utilized the \taurex\ forward model to predict the spectra of planets in primary and secondary eclipses.
In order to account for the stellar activity we made use of the observed solar XUV spectrum taken from Virtual Planetary Laboratory (VPL)
as a proxy for an active sun-like star.}
{We find large changes in mixing ratios of most chemical species in planets orbiting A-type stars that radiate strong XUV flux
inducing a very effective photodissociation.
For some species, these changes can propagate very deep into the planetary atmosphere to pressures of around 1~bar.
To observe disequilibrium chemistry we favor hot Jupiters with temperatures
$\teq=1000$~K and ultra-hot Jupiters with $\teq\approx3000$~K that also have temperature inversion in their atmospheres. 
On the other hand, disequilibrium calculations predict no noticeable changes in spectra of planets with intermediate temperatures.
We also show that stellar activity similar to the one of the modern Sun drives important changes in mixing ratio
profiles of atmospheric species. However, these changes take place at very high atmospheric altitudes and thus do not affect
predicted spectra. Finally, we estimate that the effect of disequilibrium chemistry 
in planets orbiting nearby bright stars could be robustly detected and studied with future missions with spectroscopic capabilities 
in infrared such as, e.g., JWST and ARIEL.}
{}


\maketitle

%

\section{Introduction}
Since the moment of their birth and the accretion of first atmospheres, the evolution of planets
is closely related to the evolution of their host stars through a wealth of different processes
that are usually summarized under a single definition of the star-planet interaction.
For example, it includes the tidal interaction between the star and the planet \citep{2010ApJ...723.1703D}
with two planets detected close to their tidal disruption limit \citep{2016MNRAS.458.4025D,2014MNRAS.440.1470B}.
If the planet is very close to its host star, the interaction between the stellar and planetary magnetic fields
can induce active regions on the surface of the host \citep{2019NatAs.tmp..408C}.
In rocky planets, strong gravitational tides and/or induction heating
can significantly enhance {volcanic} activity and hence outgassing which can create dense Venus-like atmospheres \citep{2017NatAs...1..878K}.
Finally, stellar radiation shapes the temperature distribution in planetary atmospheres, and stellar activity produces strong non-thermal radiation
(X-ray and UV~--~XUV) and energetic particle flux (coronal mass ejections~--~CME) that all drive atmospheric chemistry out of equilibrium and eventually 
control processes of atmospheric erosion \citep{2019A&A...624L..10J,2019MNRAS.487.4208D}. 
Among all these processes, the stellar radiation is of particular interest
because it acts on the atmosphere of every planet and its impact on, e.g., atmospheric composition
could directly be studied by analyzing the spectra of the planetary atmosphere.

Giant planets at short orbital distances {(hot Jupiters (HJs) and smaller Neptune-size planets)}
receive strong radiation from their host stars which raises atmospheric
temperatures to thousand and, in some extreme cases of so called ultra-hot Jupiters (UHJ), to even $\teq$$=$4000~K 
for the famous KELT-9b planet \citep{2018NatAs...2..714Y}.
This leads to an intense photo-dissociation of molecules at high altitudes in their atmospheres, so that the expected concentrations
of molecular and atomic species may start to {strongly deviate} from their equilibrium values.
This deviation becomes highly noticeable when, additionally to photochemistry, the transport processes (e.g., molecular diffusion) are taken into account.
HJs are the best studied class of exoplanets to date because they are easily to observe thanks to their {short orbital periods,} large sizes, and
hot atmospheric temperatures (which makes the planet-to-star flux contrast relatively large). {This makes HJs the most promising targets to study
atmospheric chemistry in detail with current and future instruments and missions.}

{Two main disequilibrium processes modify abundances within exoplanetary atmospheres.
The first one is photochemistry which includes dissociation and ionization of atmospheric constitutes by stellar XUV radiation.
The second is vertical mixing which regulates how the atmospheric species segregate within the atmosphere under 
the action of competing processes of molecular diffusion and turbulent mixing.}
The consequences of disequilibrium chemistry on the atmospheric composition of different kind of exoplanets have
been explored by numerous groups in the past.
{Early studies specifically addressed questions of photochemically driven production of atomic species 
and thermospheric escape in HJ atmospheres \citep{2003ApJ...596L.247L,2004Icar..170..167Y,2007ApJ...661..515K,2007P&SS...55.1426G}
as well as pathway analysis of different photochemical products \citep[e.g.,][]{2009ApJ...701L..20Z,2010ApJ...717..496L}.
These studies led to the development of new models with improved photochemical and kinetic calculations.
These models were used to study disequilibrium processes in objects of different masses and temperatures:
in terrestrial exoplanet atmospheres using benchmark cases of Mars and Earth \citep[e.g.,][]{2012ApJ...761..166H}, 
super-Earths \citep[GJ~1214~b,][]{2012ApJ...745....3M},
hot Neptines \citep[GJ~436~b,][]{2011ApJ...738...32L}, and HJs \citep{2014ApJ...780..166M,2012ApJ...745...77K}.
In-depth analysis of disequilibrium effects (including realistic treatment of activity driven stellar XUV radiation)
and validation of chemical networks was performed for the two benchmark cases of HD~209458~b and HD~189733~b, both planets having close
equilibrium temperatures ($\teq$$=$1200~K and $\teq$$=$1500~K, respectively) 
but orbiting stars of different types and ages \citep{2011ApJ...737...15M,2012A&A...546A..43V}.
{In UHJs, such as KELT-9, high temperatures and strong irradiation causes large amounts of atoms and ions
to be present in upper atmospheric layers,  as was predicted by \citet{2018ApJ...863..183K} and later detected 
using high-resolution spectroscopic observations by \citet{2019A&A...627A.165H}. However, these two studies did not attempt
detailed prediction of the impact of disequilibrium processes on the spectra of KELT-9.}
Finally, a possibility to study disequilibrium chemistry in atmospheres of three planets 
(HD~189733~b, WASP-80~b, and GJ~436~b, all having $\teq$$<$1200~K) with future space missions was carried out in \citet{2018ApJ...853..138B},
concluding that, e.g., with James Webb Space Telescope (JWST) it will be possible to robustly constrain the difference between equilibrium and disequilibrium chemistry.}
Analyzing chemistry in HJs provides an important test for existing kinetic models {as they} include many poorly
known parameters (turbulent diffusion coefficients, photodissociation cross-section, reaction rate coefficients, choice of the kinetic network, etc.). 
When the atmosphere of a planet is driven away from its chemical equilibrium, 
the changes in molecular concentrations start to affect the temperature structure of the atmosphere via the changes in local
opacity. That is, the initially assumed temperature structure may change if the changes in opacity are strong enough.
Ideally, this non-linear response of the atmospheric structure to the disequilibrium processes
must be taken into account in a self-consistent way by solving for the temperature structure and number densities simultaneously.
{This enormously complicates the problem as solving for the temperature structure requires 
the knowledge of frequency integrated properties of the radiation field at every point in the planetary atmosphere.}
As a result, many works addressed only the effect of photochemistry and kinetics on the changes in mixing ratios and observed characteristic
of some benchmark HJs \citep[see, e.g.,][]{2019MNRAS.487.2242H,2012A&A...546A..43V,2011ApJ...737...15M}.
Recently, \citet{2019ApJ...883..194M} investigated the effect of disequilibrium chemistry
on the observed properties of planetary atmospheres using an extensive grid of models that covered a very large range of planetary and stellar
parameters. {However, they did not specifically address the effect of stellar activity as a function of stellar age
(limiting only to the XUV flux of the present Sun) and of inverted temperature profiles expected for UHJs.}

Self-consistent kinetic-structure models of planetary atmospheres remain a challenging task, but the first models have already been successfully developed in the past. 
For instance, \citet{2016A&A...594A..69D} used the \atmo\ code that can simultaneously solve for kinetic and radiative-convective equilibrium
to study the temperature structure and emission spectra of HJs. They concluded that by not accounting for the radiative-convective equilibrium
can lead to the overestimation of the effect of disequilibrium chemistry on the observed spectra. The authors highlighted the importance of self-consistent models
because, although the emission spectra did not show large changes between equilibrium and disequilibrium calculations, 
the mixing ratios and atmospheric temperature were both highly affected by photodissociation
and molecular diffusion processes.

In this work we aim at a systematic analysis of disequilibrium chemistry in an atmosphere of a Jupiter size exoplanet
that orbits stars of different spectral types. Note that we do not attempt to compute self-consistent
kinetic models similar to \citet{2016A&A...594A..69D}. Instead, here we make a first step
towards extending our study of disequilibrium chemistry considering different stellar spectral types and we make predictions for both emission and transmission observations.
Moreover, studying the impact of disequilibrium chemistry, we additionally 
investigate the impact of stellar activity on atmospheres of HJs orbiting young sun-like stars that maintain
high level of XUV radiation during early stages of their evolution.

\section{Methods}\label{sec:methods}

\subsection{Atmospheric pressure-temperature profiles}

We utilized the \helios\footnote{\tt https://github.com/exoclime/HELIOS}
code to compute temperature structure of atmospheres using self-consistent radiative-convective iteration \citep{2017AJ....153...56M}.
The typical inputs for the \helios\ are the parameters of the planet (mass, radius, atmospheric abundances, semi-major axis) and its
host star (radius, effective temperature) and we used the black body approximation for the stellar spectrum.
Our calculation setup included 100 atmospheric layers distributed logarithmically between $10^{-9}$ bar and $100$ bar and we assumed
global redistribution of the heat across day and night sides of the planet.

Modeling atmospheres of UHJs deserves additional attention. Intense XUV radiation of hosts stars drives
very efficient photo-dissociation of molecules which then enhances concentrations of atomic species and their ions
in the atmospheres of these planets \citep{2019A&A...628A...9C}. 
The amount of these metals could be so large that they begin to play a dominant role in shaping
the P-T structure by further absorbing XUV stellar flux. Most of the absorbed energy goes into the kinetic energy 
of the ionized atoms and electrons. The result of this process is the sudden temperature raise detected in corresponding atmospheric layers 
\citep{2019AJ....157..170M,2018ApJ...855L..30A}.
It was then understood that a proper modeling of UHJ atmospheres requires inclusion of opacity due to metals and their ions.
Unfortunately, in its current stage the opacity tables included in the public version of the \helios\ code do not include continuum and line opacity
due to atoms. We therefore considered two approaches.
First, because our analysis has an exploratory nature and does not involve direct comparisons with real data, we still used \helios\ to compute
T-P profiles for most irradiated planets. Second, we considered analytical T-P profiles after \citet{2014A&A...562A.133P} 
in order to explore the effect of {temperature inversion} on the chemical profiles and  predicted spectra of UHJ.
{Since we are interested in studying general changes in predicted spectra of planets with temperature inversion 
in their atmospheres, we have chosen a parameterized T-P model with an inversion at the mbar level. 
Note that we do not compare the resulting synthetic spectra obtained by using temperature profiles from HELIOS 
with those from parameterized models. The temperature inversion can be obtained either with \citet{2014A&A...562A.133P} or, e.g., with another
commonly used parameterization after \citet{2010A&A...520A..27G}.
An extensive use of these parameterized models can be found in, e.g., \citet{2015A&A...574A..35P} and \citet{2015A&A...577A..33V}.}

\begin{figure}
\centerline{
\includegraphics[width=\hsize]{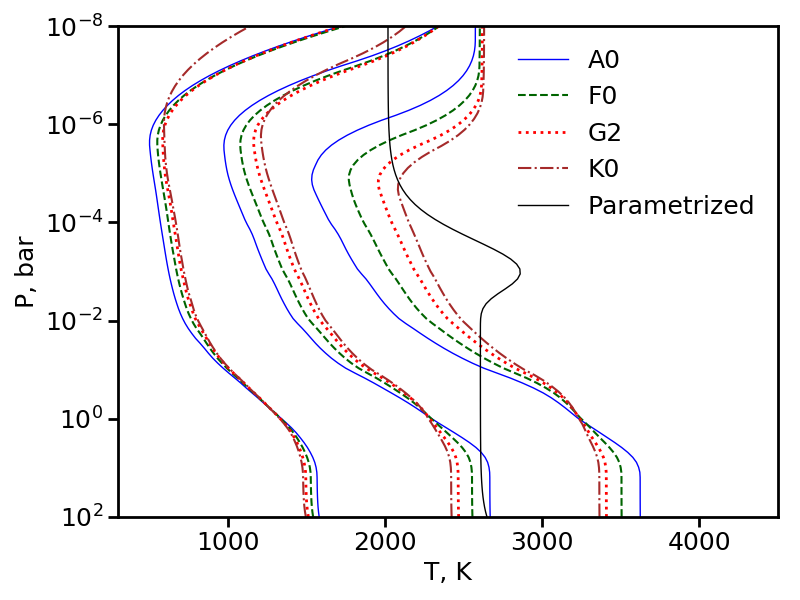}
}
\caption{Temperature profiles of a Jupiter size planet around
three types of host stars (A0~--~solid blue line, F0~--~dashed green line, G2~--~dotted red line, K0~--~dash-dotted brown line)
calculated with \helios\ and grouped according to the equilibrium temperatures of 1000~K (left), 2000~K (middle), and 3000~K (right), respectively.
The parameterized profile after \citet{2014A&A...562A.133P} with temperature inversion in upper atmospheric layers is shown with full black line.
}
\label{fig:t1t2t3}
\end{figure}

{Our Fig.~\ref{fig:t1t2t3} illustrates examples of temperature profiles corresponding to $\teq$=1000~K, 2000~K, 3000~K, 
and inverted $\teq$=3000~K, respectively, for a Jupiter size planet orbiting stars of different spectral types.
Note the difference between temperature profiles introduced by stellar types. This difference is obviously absent in case of the parameterized profile.
The equilibrium temperatures were estimated assuming planetary albedo $\alpha$$=$0.03 \citep{2000ApJ...538..885S}.}

\subsection{Chemical kinetic model}
The vertical distribution of neutral atmospheric species in the atmosphere is obtained by solving the coupled kinetic
equations for species containing C, H, O, and N using the \vulcan\ code\footnote{\tt https://github.com/exoclime/VULCAN} \citep{2017ApJS..228...20T}.
{All absorption and photodissociation cross sections are from Leiden Observatory database\footnote{\tt http://home.strw.leidenuniv.nl/\~ewine/photo}
\citep[see also][]{2017A&A...602A.105H}, 
except NO$_{\rm 3}$, HNO$_{\rm 3}$, HNO$_{\rm 2}$, N$_{\rm 2}$O, and NO$_{\rm 2}$ that are from 
\citet{HUEBNER201511,1992Ap&SS.195....1H,1979STIN...8024243H} where NO$_{\rm 2}$ combines $\rm NO/O(^1D)$ and NO/O 
to obtain the total absorption cross section. N$_{\rm 2}$H$_{\rm 4}$ is from MPI-Mainz UV/VIS Spectral Atlas\footnote{\tt http://satellite.mpic.de/spectral\_atlas/index.html}.}
We assumed that the metallicity of the exoplanetary atmosphere is solar \citep{2009LanB...4B..712L}, that is  $\rm C/H=2.45\times 10^{-4}$, 
$\rm O/H=4.57\times10^{-4}$, and $\rm N/H=6.03\times10^{-5}$. This renders a C-to-O ratio to 0.54. The {chemical kinetic} equations are solved between 100 and
$10^{-8}$ bar with boundary conditions of zero flux at both atmospheric levels. The thermal profile T(p) and the metallicity 
are the only inputs needed by \fastchem\footnote{\tt https://github.com/exoclime/FastChem} \citep{stock_kitzmann_patzer_sedlmayr_2018}
to obtain the thermochemical 
equilibrium abundances of the considered species. The obtained vertical profiles are then subject to photodissociation 
caused by the stellar flux, turbulent transport, and molecular diffusion with the \vulcan\ code.

\begin{figure}
\centerline{
\includegraphics[width=\hsize]{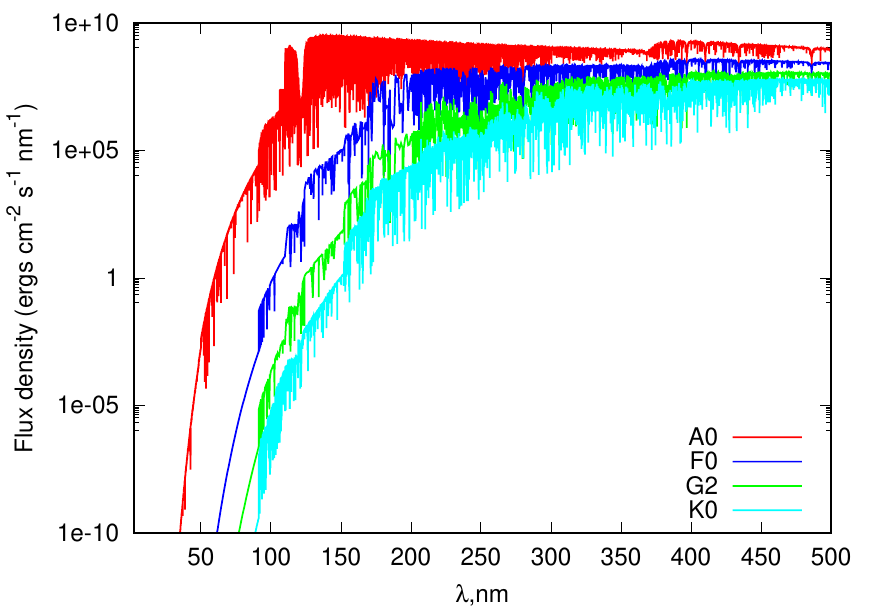}
}
\caption{Flux density, at the stellar surface, for A0, F0, G2, and K0 stars as obtained from \phoenix\ model library \citep{2013A&A...553A...6H}. 
}
\label{fig:stellar_fluxes}
\end{figure}

The stellar flux was taken from the \phoenix\ library \citep{2013A&A...553A...6H} and scaled according to the orbital distance
of the planet and the size of the host star. Each stellar type that we considered in this study (A0, F0, G2, K0) was assumed to have solar metallicity. 
Figure \ref{fig:stellar_fluxes} shows the spectral flux energy at the stellar surface at $\lambda \leqslant 500$~nm as this is the spectral
region essential for the bulk of the photodissociation processes.

{We purposely ignored M dwarfs in our study because these stars are not expected to form
many hot Jupiters \citep{2016A&A...587A..49O,2012A&A...541A..97M}. Up to now only a few Jupiter size planets were
discovered around M dwarfs \citep{2018MNRAS.475.4467B,2018arXiv181209406B,2015AJ....149..166H,2012AJ....143..111J}. 
Note that all of them have $\teq<1000$~K which is outside of the temperature range considered in this work.}

The turbulent transport is usually parameterized in 1D models by so called eddy diffusion coefficient $K_{zz}$, which is associated 
with the vertical mixing. As in \citet{2018ApJ...863..183K}, we assumed $K_{zz}=0.1 H_p v_z$, where $H_p$ refers to the atmospheric scale height, and 
$v_z$ to the vertical velocity which is approximated to be the atmospheric speed of sound $c_s$. 
This consideration turns into $K_{zz}$ ranging from $4.3\times 10^9$ to $2.4\times 10^{10}$ \dif\ for $\teq$ between 1000~K and 3000~K.
The molecular diffusion coefficients were computed following \citet{2011ApJ...737...15M,2000Icar..143..244M} 
{and they depend on temperature, total number density, and mass of the diffusing species. Other approaches
can be taken as well (e.g., Chapman and Enskog equation from \citet{poling2000properties}).}

\subsection{Predicting emission and transmission spectra of HJs}

\begin{figure}
\centerline{
\includegraphics[width=\hsize]{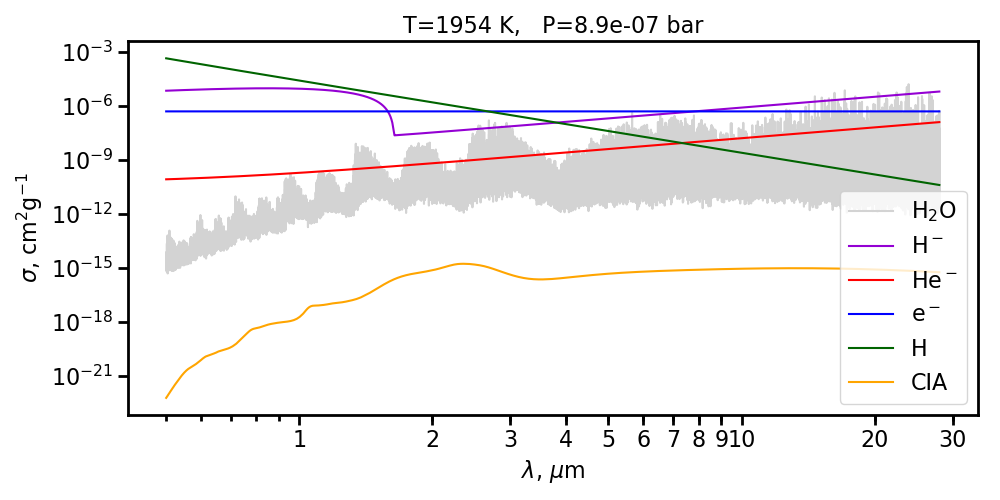}
}
\centerline{
\includegraphics[width=\hsize]{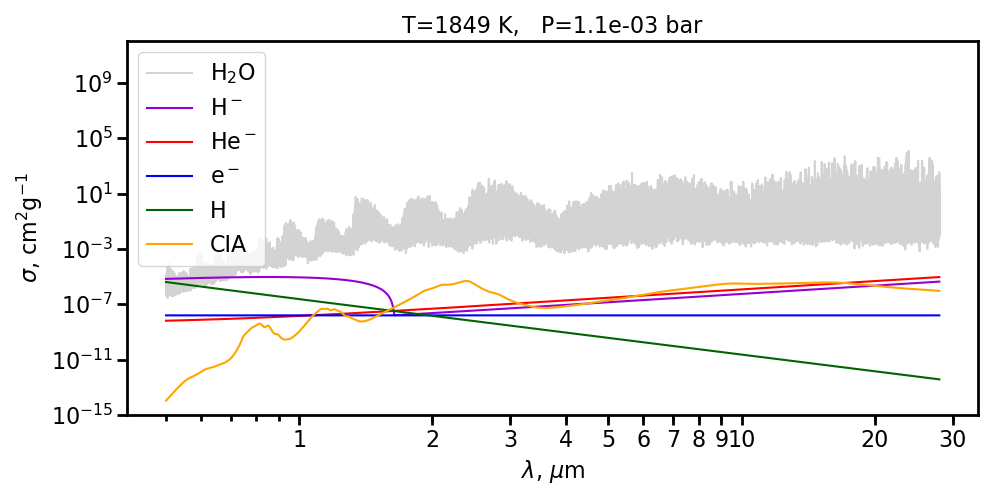}
}
\centerline{
\includegraphics[width=\hsize]{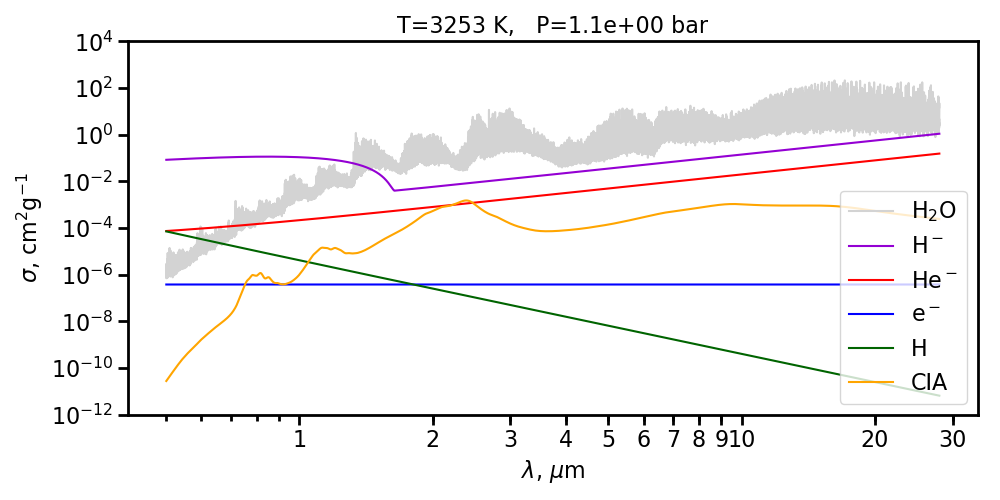}
}
\centerline{
\includegraphics[width=\hsize]{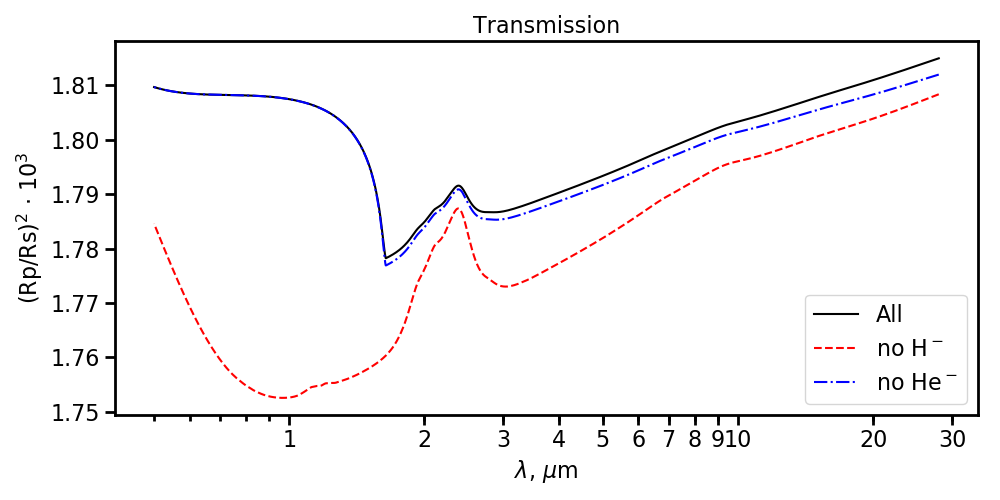}
}
\caption{Continuum opacity calculated for a HJ with $\teq$$=$3000~K at three different altitude levels (with pressure decreasing from bottom to top).
The opacity due to \hho\ is shown for comparison purpose. The bottom plot shows the transmission amplitude calculated including only
continuum opacity sources (i.e., without line opacity) and illustrates the relative contribution of \hminus\ and \heminus\ absorption.}
\label{fig:opac}
\end{figure}

In order to predict emission and transmission spectra of HJs we utilized the \taurex~(Tau Retrieval for Exoplanets) 
software package \citep{2015ApJ...802..107W,2015ApJ...813...13W}.
It uses up-to-date molecular cross-sections based on line lists provided by \textsc{ExoMol}\footnote{\tt www.exomol.com}
project \citep{2012MNRAS.425...21T} and \textsc{HITEMP} \citep{2010JQSRT.111.2139R}. We additionally used the \hcn\ line list after \citet{2006MNRAS.367..400H}
provided by the \textsc{Exoclime} project\footnote{\tt https://dev.opacity.iterativ.ch/\#/} and generated with the \heliosk\ code \citep{2015ApJ...808..182G}.
These cross-sections are pre-computed on a grid of temperatures and pressures
and are stored in binary opacity tables that are available for a number of spectral resolutions. The continuum opacity includes
Rayleigh scattering on molecules and collisionaly-induced absorption due to H$_{\rm 2}$-H$_{\rm 2}$ and H$_{\rm 2}$-He either
after \citet{2011mss..confEFC07A,2012JChPh.136d4319A} or \citet{2001JQSRT..68..235B,2002A&A...390..779B,1989ApJ...341..549B}, respectively.
We extended the public version of \taurex\footnote{\tt https://github.com/ucl-exoplanets/TauREx\_public} by incorporating
additional opacity sources essential for the atmospheres of UHJs. In particular, bound-free and free-free transitions of \hminus\ become
one of the major continuum opacity contributors for the temperatures hotter than about 2000~K. 
{The cross-sections of \hminus\ are from \citet{1988A&A...193..189J}.
We also included opacity due to free-free
transitions of \heminus\ as well as Rayleigh scattering on \ion{H}{i} atoms and Thomson scattering on free electrons.
The \heminus\ cross-sections are those originally from \citet{1968MNRAS.138..137J} using polynomial fit by \citet{1969lls..symp..435C}.
Rayleigh scattering on \ion{H}{i} is calculated after \citet{Dalgarno1962SPECTRALRO}.
All relevant numerical routines were extracted from the \llm\, stellar model atmosphere code \citep{2004A&A...428..993S}.
At the temperatures of HJs the \hminus\ is the dominant continuum opacity source that impacts the 
observed spectra of these planets \citep{2018ApJ...855L..30A}. However, at millibar pressures the \heminus\ opacity 
can become comparable or even stronger than that of \hminus\ for wavelengths longer than 1.6~\mum, as shown
on Fig.~\ref{fig:opac} (third panel from the bottom) where we display examples of continuum opacity coefficient
at different altitudes in the atmosphere of a HJ with $\teq$$=$3000~K. 
At even smaller pressures, electron scattering and Rayleigh scattering on \ion{H}{i} atoms
also become important contributors to the continuum opacity at particular wavelengths (top panel on Fig.~\ref{fig:opac}). 
However, the contribution of \ion{H}{i} and \electron\ on the transmission spectra is marginal 
because their opacity is strong only at low pressures that are hardly probed by transmission spectroscopy. 
Thus, among all continuum opacity sources, only \hminus\ and \heminus\ significantly contribute to the predicted amplitude of the transmission spectra,
as shown on the bottom plot of Fig.~\ref{fig:opac}. When both continuum and line opacity are included, 
the effect of \heminus\ on predicted spectra is diluted by a much stronger opacity in molecular lines
while \hminus\ contribution is still significant.
Nevertheless, as can be seen from the second plot (from bottom) on Fig.~\ref{fig:opac}, in optically thick layers the \heminus\ opacity 
could still be stronger than, e.g., collision-induced absorption due to H$_{\rm 2}$-H$_{\rm 2}$ and H$_{\rm 2}$-He. 
We thus conclude that accurate calculation of atmospheric opacity requires \heminus, \ion{H}{i},
and \electron\ opacity included especially at low pressures, similar to how it is done in modern stellar atmosphere codes. 
However, observed transmission and emission spectra of HJs are hardly affected by these three opacity sources. 
Finally, we updated \helios\ with \heminus\ and \electron\ opacity 
(original version of \helios\ already includes \ion{H}{i} Rayleigh scattering)
and found out that this has only little impact on the atmospheric temperature structure (with a modification in local temperature
of at most $\Delta$T$\approx$10~K) and thus can be ignored.}

\subsection{Stellar activity} \label{stellar_activity}
Dynamos in stars with outer convective envelopes can generate
strong magnetic fields that interact with stellar convection and eventually dissipate their energy 
in magnetic re-connection events. The latter heat up the regions above the stellar photospheres and create chromospheres and coronae.
High temperature in those atmospheric regions enhances stellar XUV flux.

During their life on the main sequence stars with convective envelopes undergo two major regimes of their activity evolution. 
First, when stars reach the main-sequence, they generally rotate very fast and their
non-thermal emission is maximal. This is the regime of activity saturation
because the amount of XUV flux does not change with time as rotation rate keeps decreasing
due to the magnetic braking. The level of the magnetic activity in this regime is determined, to a very good approximation,
by the thickness of the convective zone which sets the amount of magnetic energy the star can generate \citep{2009Natur.457..167C}.
As stars evolve they keep spinning down and after some time reach a critical rotation period above which the activity starts to dissaturate
and begin to decline as rotation rate of stars decreases further (non-saturated regime, see, e.g., \citet{2014ApJ...794..144R,2003A&A...397..147P,1984ApJ...279..763N}).
For instance, the Sun was more active in the past and its XUV flux was stronger which potentially drove an enhanced erosion of atmospheres
of the Earth and other planets during first several hundred million years after the formation of the solar system \citep{2019A&A...624L..10J,2015A&A...577L...3T,2005ApJ...622..680R,1997ApJ...483..947G}.
Even now, at its relatively quiet state, the Sun produces enhanced XUV emission 
{which cannot be predicted by standard photospheric models.}

Because most molecular photodissiociation cross-sections are found in the short wavelength domain,
it becomes evident that the accurate knowledge of XUV radiation is essential when studying atmospheric chemistry of planets orbiting stars with convective envelopes. 
Therefore, we extended our investigation of the G2 case by additionally accounting for the stellar activity.
Note that some semi-empirical models of stellar chromospheres have been constructed in the past to mimic the amount of non-thermal radiation
from convective stars \citep[e.g.][]{2016ApJ...830..154F,1981ApJS...45..635V}. However, calculation
of the outgoing radiation from these models requires complicated approaches including non-equilibrium effects on atoms and ions.
Instead, we specifically have chosen the case of G2 star in our stellar sample because 
it resembles the properties (temperature and radii) of the Sun \citep{2003AsBio...3..689S}.
Thus we used the observed solar radiation to address the impact of stellar activity on the atmospheric spectra of our test planet that orbits G2 star
{using Sun as a proxy of XUV radiation}.

\begin{figure}
\centerline{
\includegraphics[width=\hsize]{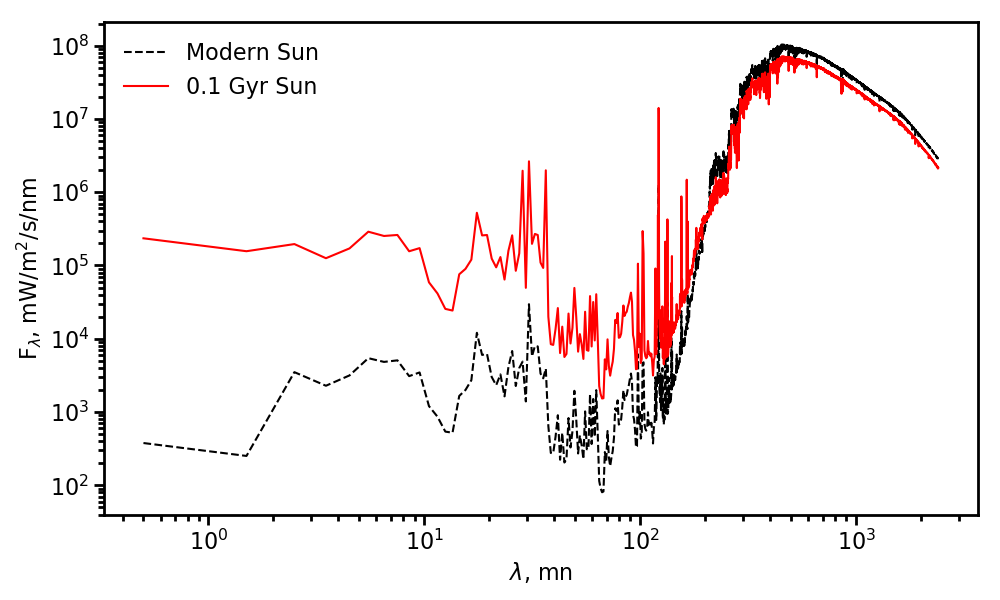}
}
\caption{Energy distribution of the Modern (dashed line) and 0.1~Gyr young Sun (solid line) at the top of the Earth atmosphere
according to \citet{2012ApJ...757...95C}.}
\label{fig:vpl-modern-young-sun}
\end{figure}

As a next step we studied the effect of an increased stellar activity for the G2 star when it was 100~Myr young.
We used estimates by \citet{2012ApJ...757...95C} who investigated the evolution of the solar flux in time
{following \citet{2005ApJ...622..680R}}. The corresponding
routines were taken from the Virtual Planetary Laboratory webpage\footnote{\tt http://depts.washington.edu/naivpl/content/models/solarflux}.
According to {the suggested scaling relations}, a 100~Myr young Sun 
would had had high-energy emissions about 100 times larger than that of the modern Sun, whereas it was less luminous and smaller
($\teff$($t$$=$4.5~Gyr)$=$5778~K; $\teff$($t$$=$0.1~Gyr)$=$5650~K, $R$($t$$=$0.1~Gyr)$=$0.876$R$($t$$=$4.5~Gyr)).
The adopted radiation of the young and modern Sun at {1 AU} are shown in Fig.~\ref{fig:vpl-modern-young-sun}.

\section{Results}

In the simulations that we present below we consider a Jupiter size planet having three different equilibrium temperatures
$\teq$ of 1000~K, 2000~K, and 3000~K, respectively. These temperatures correspond to typical conditions found in hot 
($\teq$$\leqslant$2000~K) and ultra-hot ($\teq$$>$2000~K) {Jupiters}
(note that the classification of ultra-hot {Jupiters} in terms of their temperatures is a matter of debate).
Each of this $\teq$ can be reached at different distances from the parent star depending on the spectral type of the later.
We chose four types of stars: A0 ($\teff$$=$10800~K, $R$$=$2.5$\Rsun$), F0 ($\teff$$=$7200~K, $R$$=$1.3$\Rsun$), 
G2 ($\teff$$=$5800~K, $R$$=$1.0$\Rsun$), and K0 ($\teff$$=$5200~K, $R$$=$0.85$\Rsun$)
For the UHJ atmospheres, i.e. for planets with $\teq$$=$3000~K, we additionally considered parameterized temperature profile
with the temperature inversion at high altitudes. Throughout the paper we call models that corresponds to four different 
temperatures as T1, T2, T3, and T3-inverted, respectively (see Fig.~\ref{fig:t1t2t3}).
Below we investigate the impact of stellar types on atmospheric chemistry and spectra as predicted by thermochemical equilibrium (EQ) 
and photo-kinetic disequilibrium (DQ) models.

\begin{figure*}[!ht]
\begin{minipage}{0.60\hsize}
\centerline{
\includegraphics[width=\hsize]{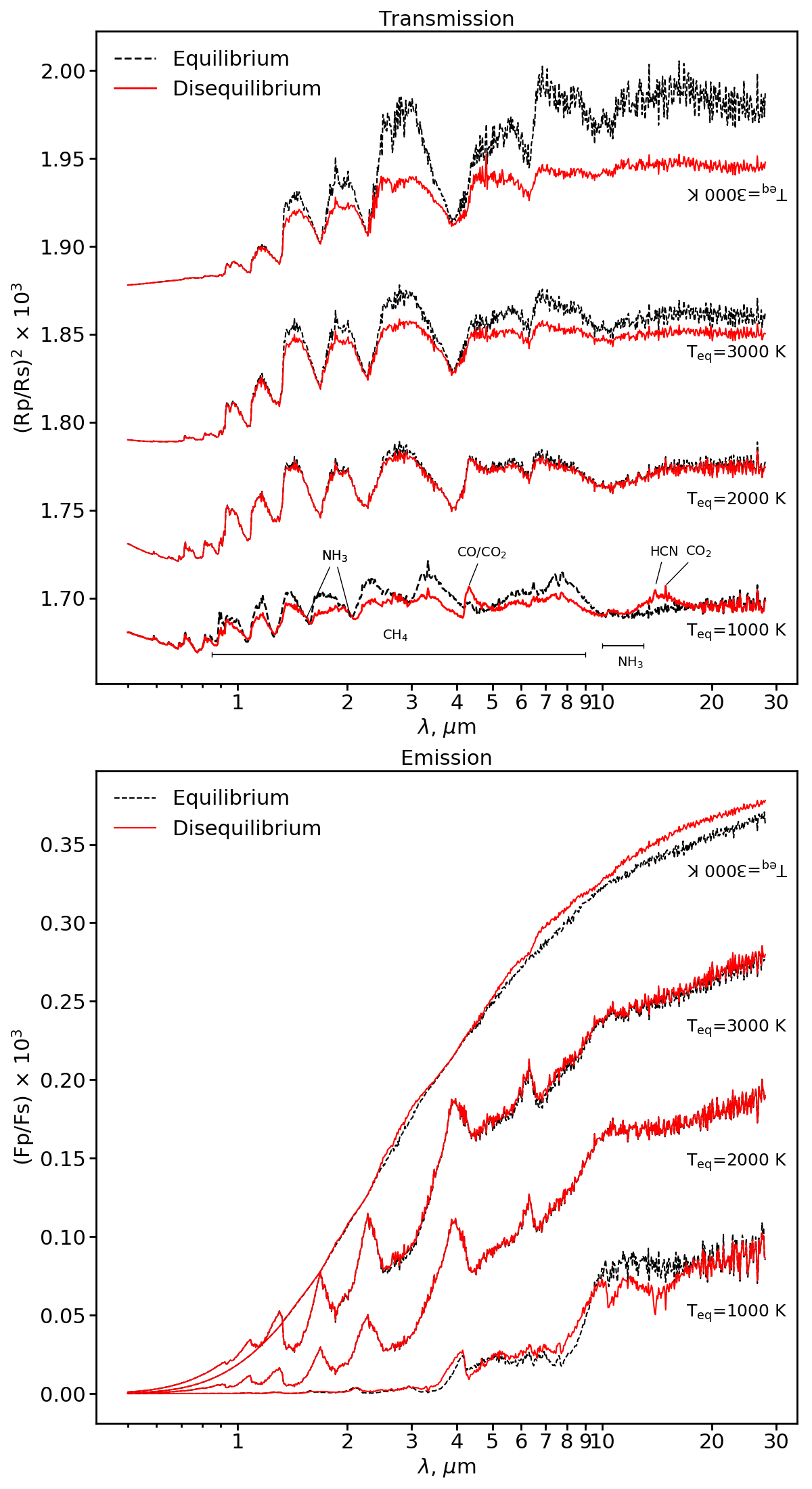}
}
\end{minipage}
\begin{minipage}{0.39\hsize}
\includegraphics[width=\hsize]{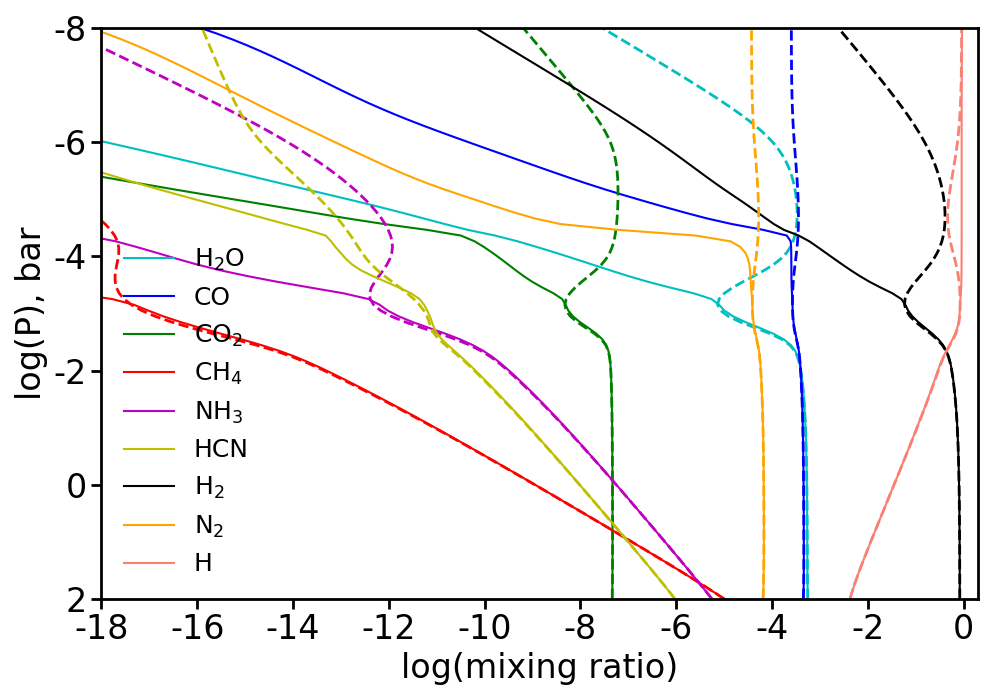}
\includegraphics[width=\hsize]{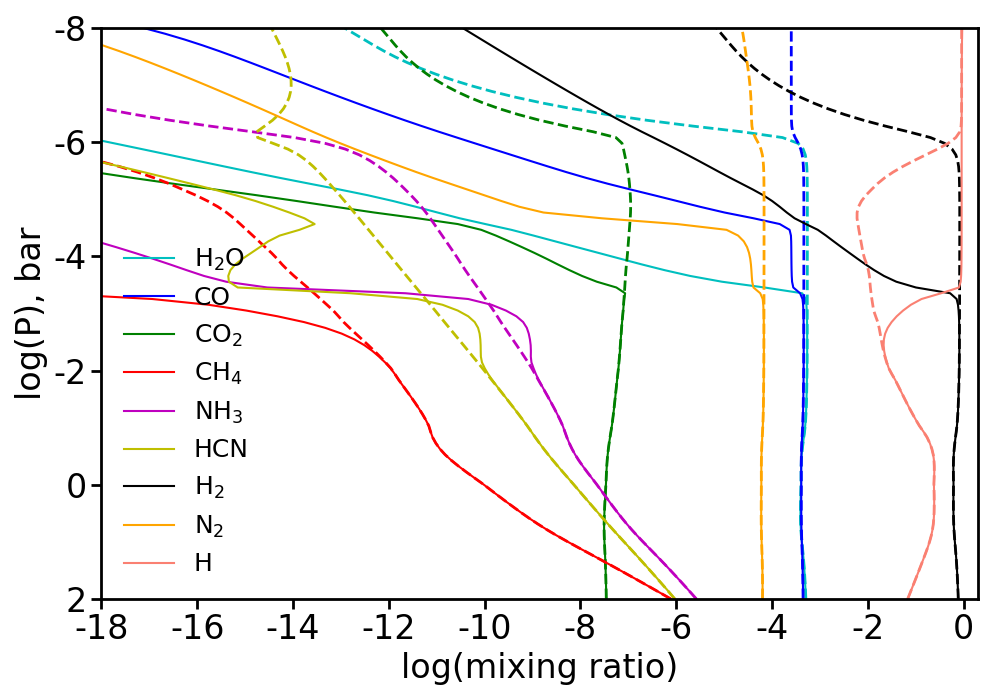}
\includegraphics[width=\hsize]{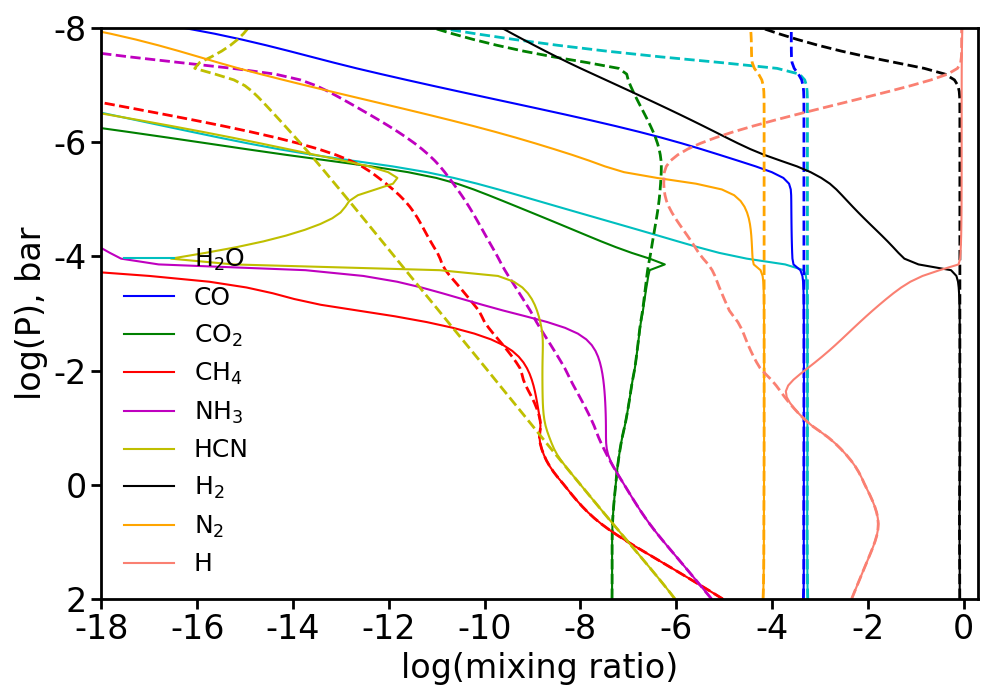}
\includegraphics[width=\hsize]{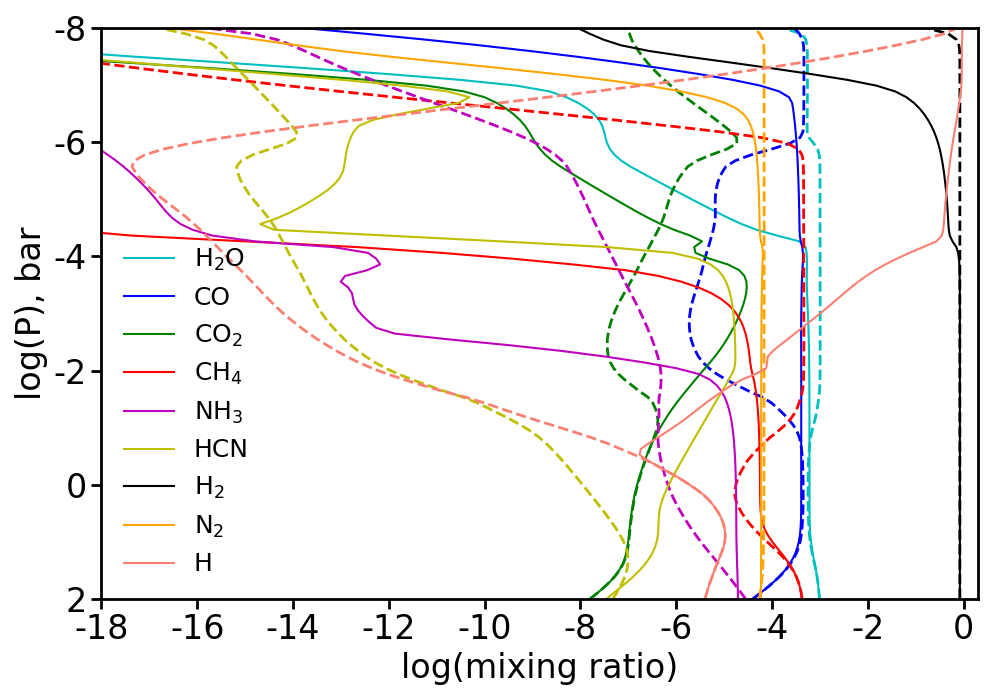}
\end{minipage}
\caption{Predicted transmission (top left) and emission  (bottom left) spectra calculated for the Jupiter size planet around an A0 star.
From bottom to top shown are the spectra predicted from the four different atmospheric structures (see text for details). 
The spectra were binned with the spectral resolution $R=\lambda/\Delta\lambda=200$.
The transmission spectra were shifted vertically for better representation.
The mixing ratios corresponding to the mentioned above atmospheric structures are shown on the second column of the figure
(from bottom to top). In all plots, the spectra and mixing ratios calculated from the equilibrium and non-equilibrium chemistry
are shown with dashed and solid lines, respectively. Disequilibrium molecules responsible for absorption bands in the emission spectrum
for the T1 case are explicitly labeled on the top left plot.}
\label{fig:sp-vmr-a0}
\end{figure*}

\begin{figure*}
\begin{minipage}{0.60\hsize}
\centerline{
\includegraphics[width=\hsize]{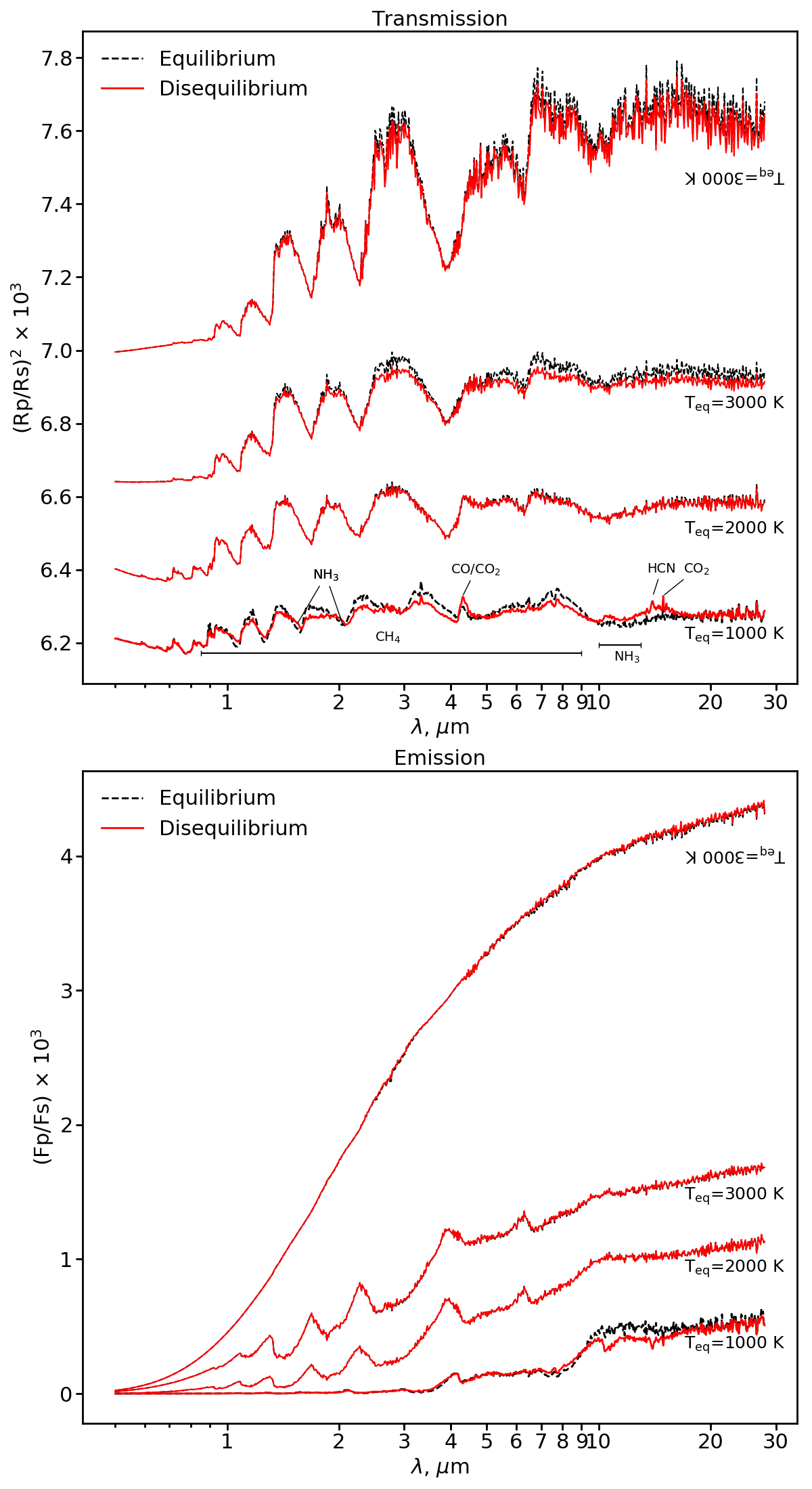}
}
\end{minipage}
\begin{minipage}{0.39\hsize}
\includegraphics[width=\hsize]{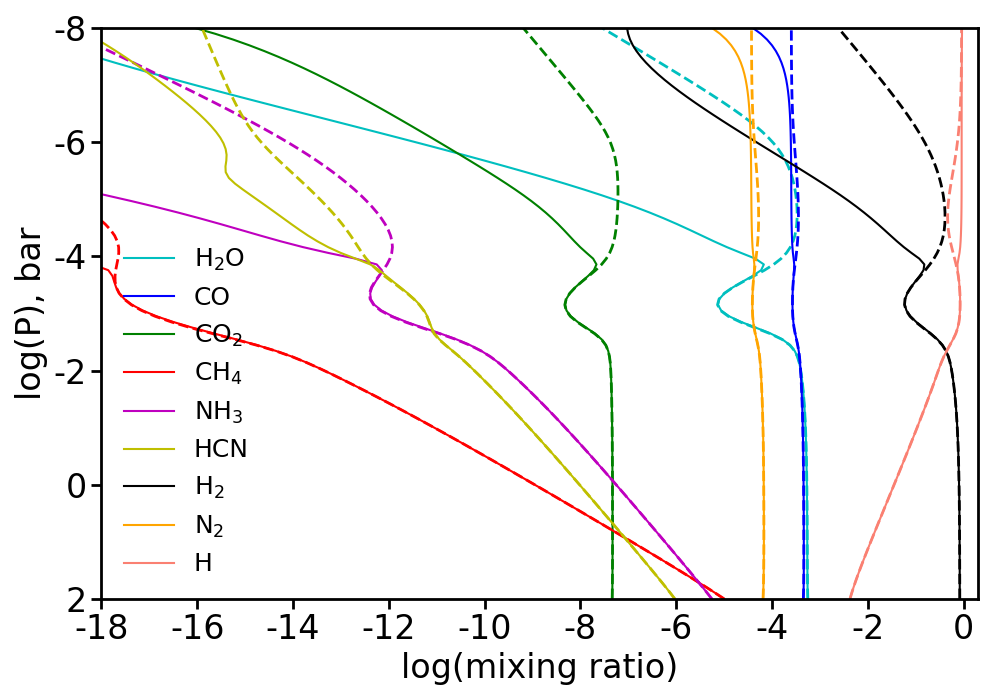}
\includegraphics[width=\hsize]{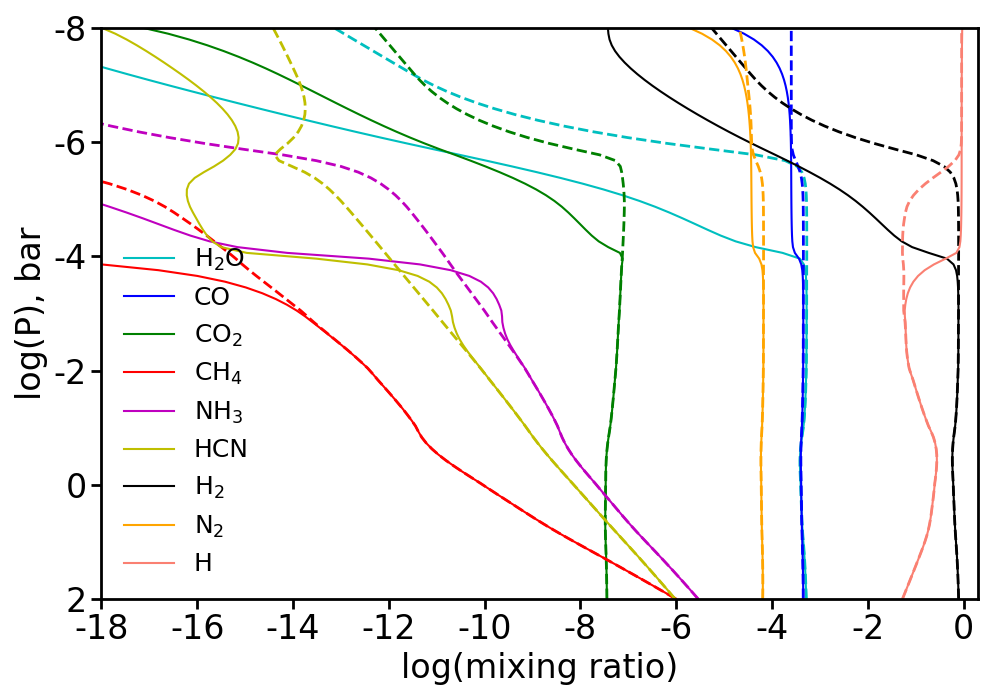}
\includegraphics[width=\hsize]{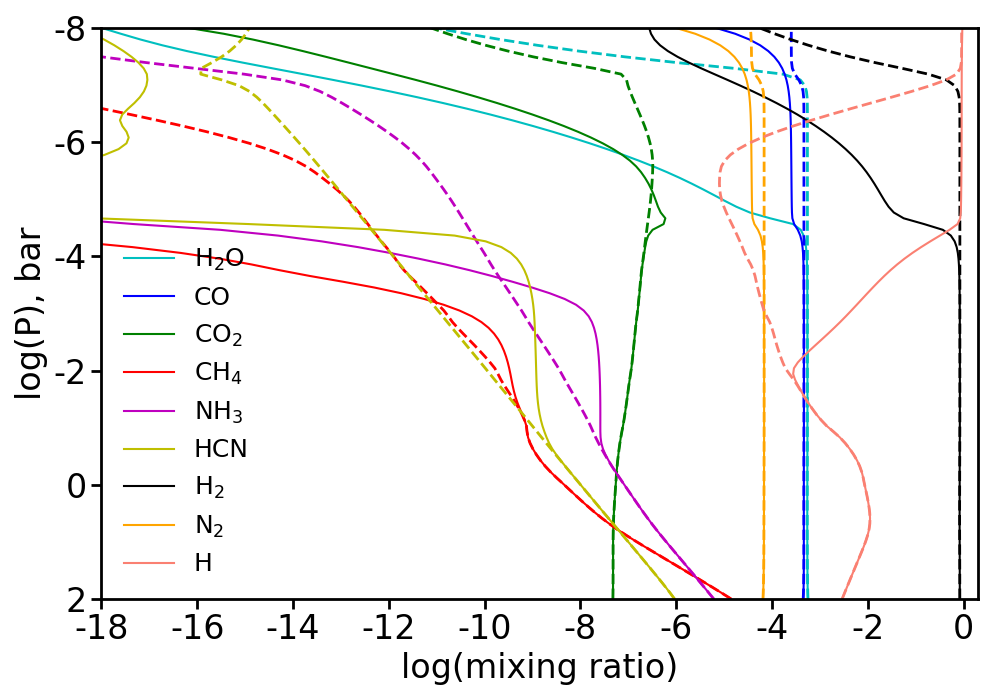}
\includegraphics[width=\hsize]{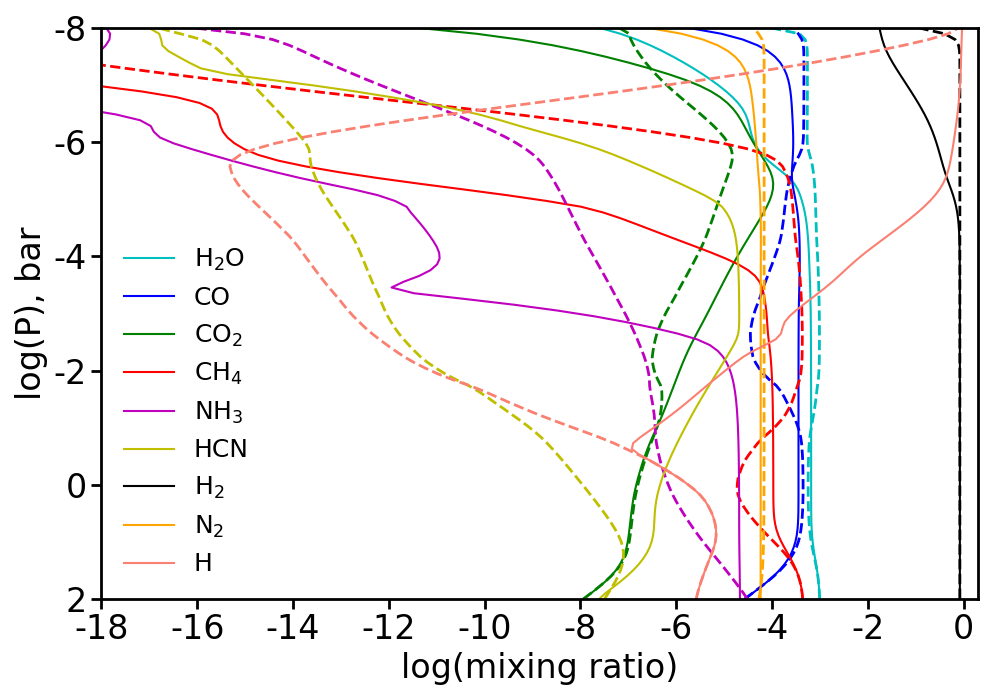}
\end{minipage}
\caption{Same as on Fig.~\ref{fig:sp-vmr-a0}, but for the planet around an F0 star.}
\label{fig:sp-vmr-f0}
\end{figure*}

\begin{figure*}
\begin{minipage}{0.60\hsize}
\centerline{
\includegraphics[width=\hsize]{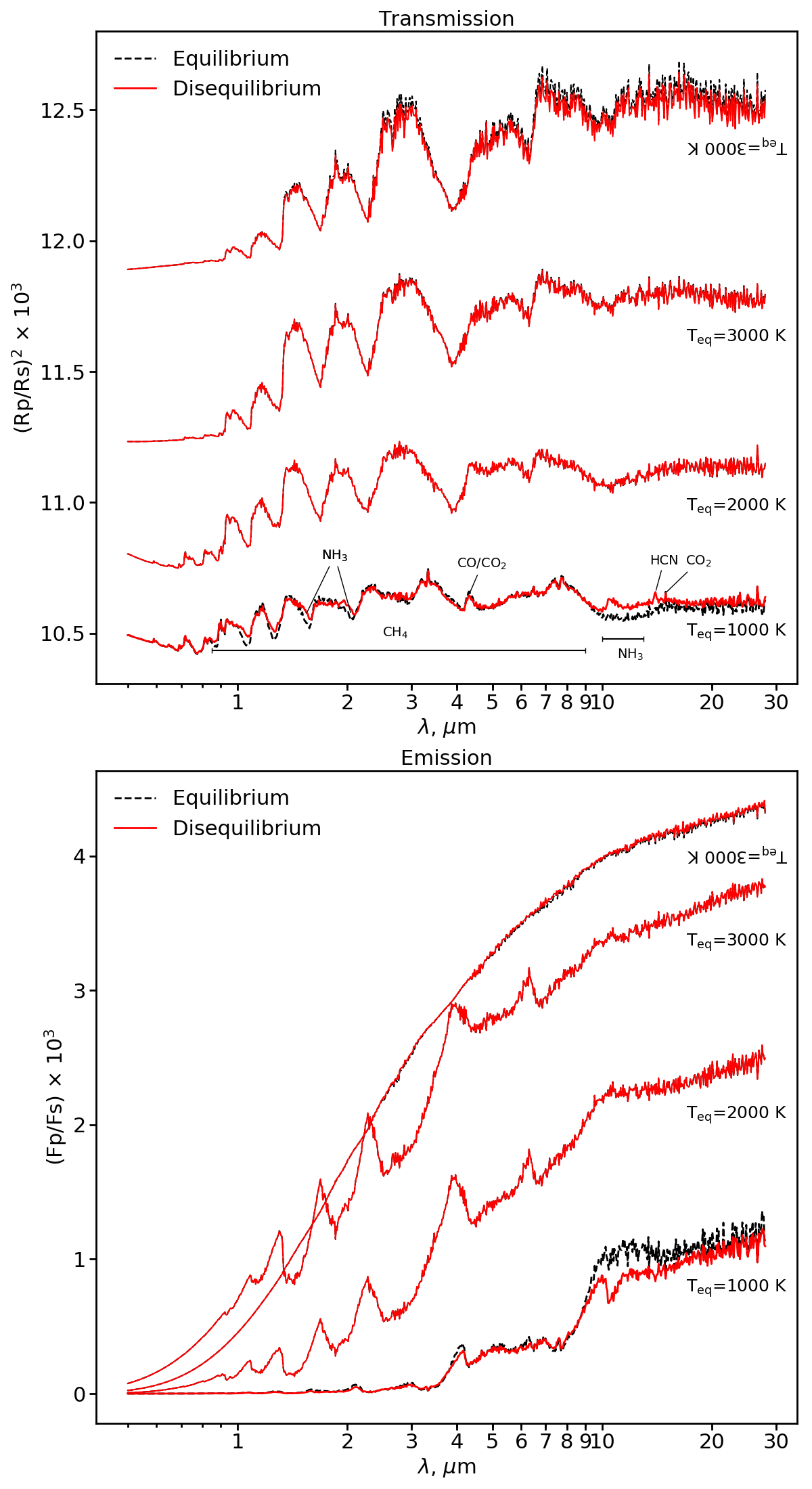}
}
\end{minipage}
\begin{minipage}{0.39\hsize}
\includegraphics[width=\hsize]{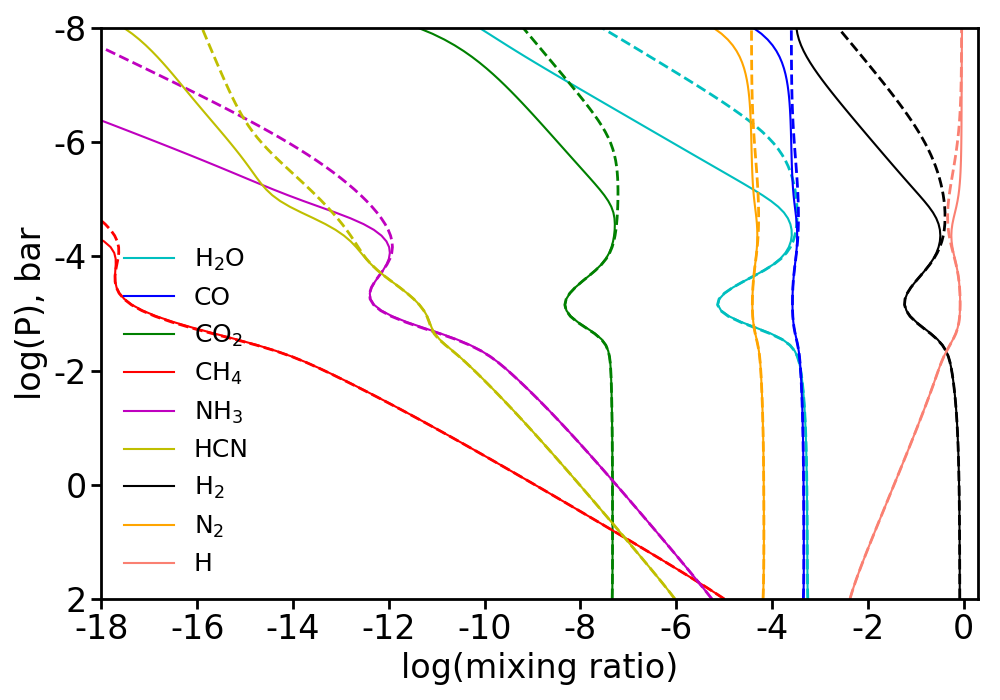}
\includegraphics[width=\hsize]{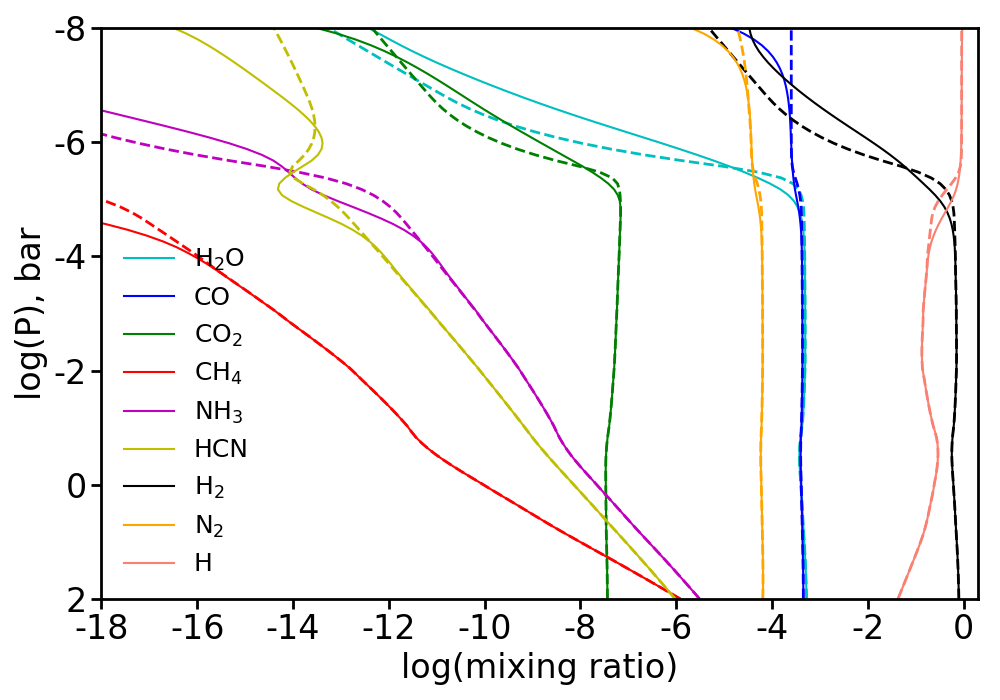}
\includegraphics[width=\hsize]{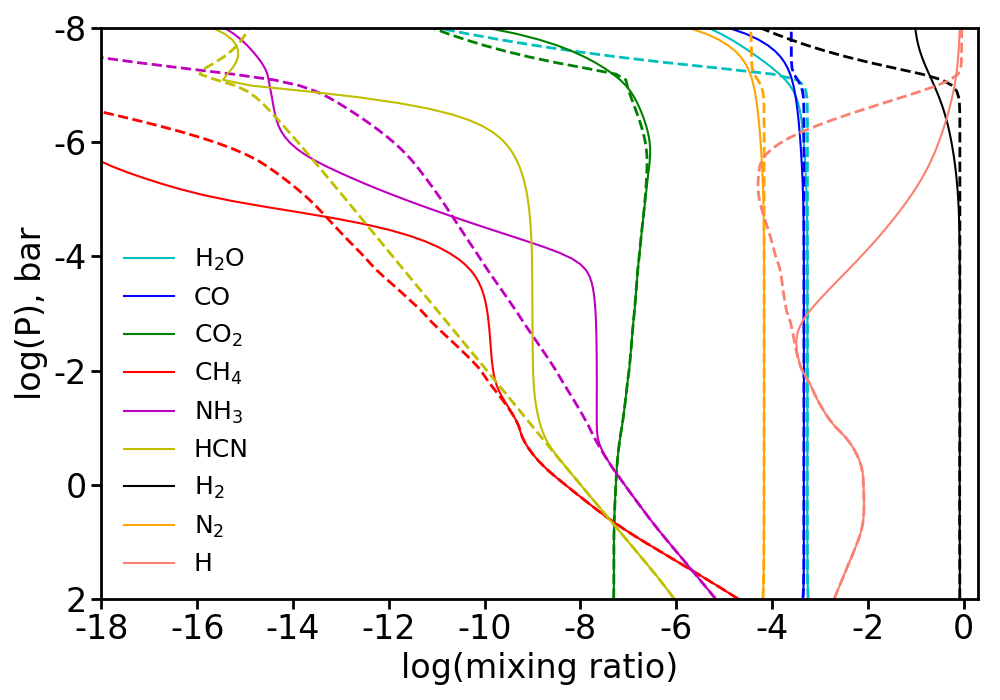}
\includegraphics[width=\hsize]{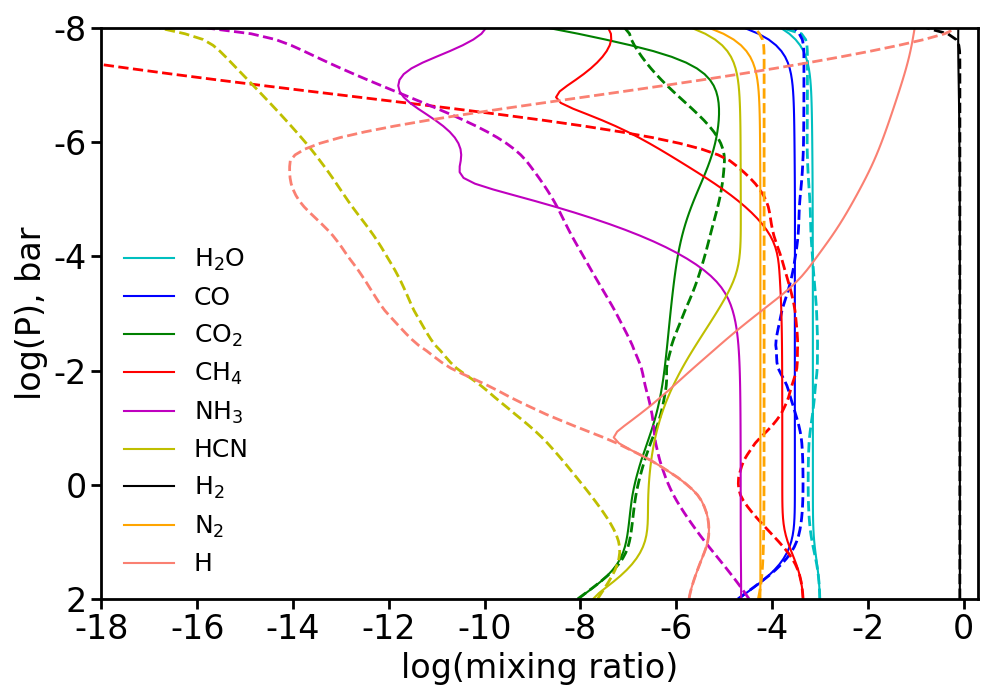}
\end{minipage}
\caption{Same as on Fig.~\ref{fig:sp-vmr-a0}, but for the planet around a G2 star.}
\label{fig:sp-vmr-g2}
\end{figure*}

\begin{figure*}
\begin{minipage}{0.60\hsize}
\centerline{
\includegraphics[width=\hsize]{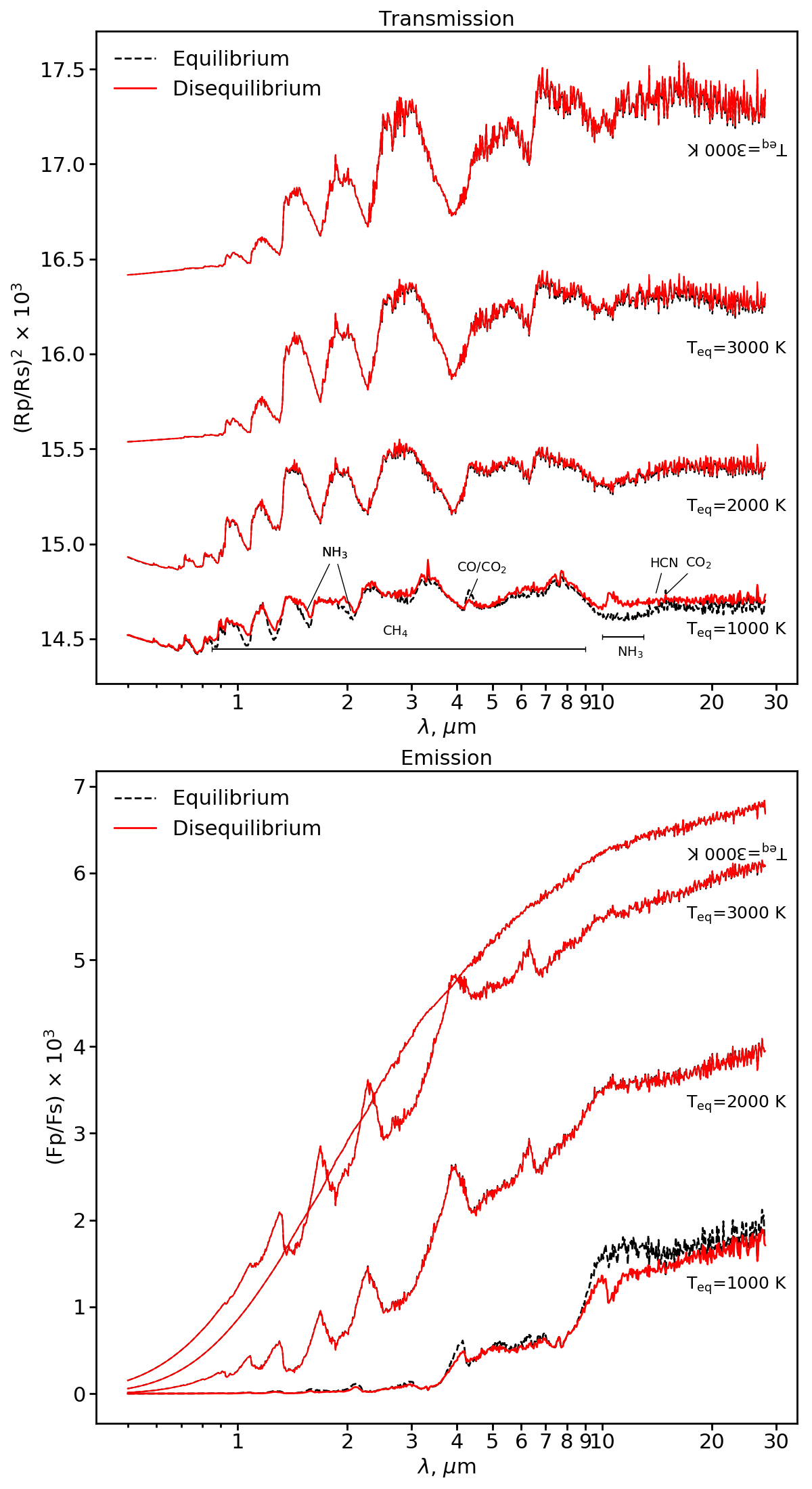}
}
\end{minipage}
\begin{minipage}{0.39\hsize}
\includegraphics[width=\hsize]{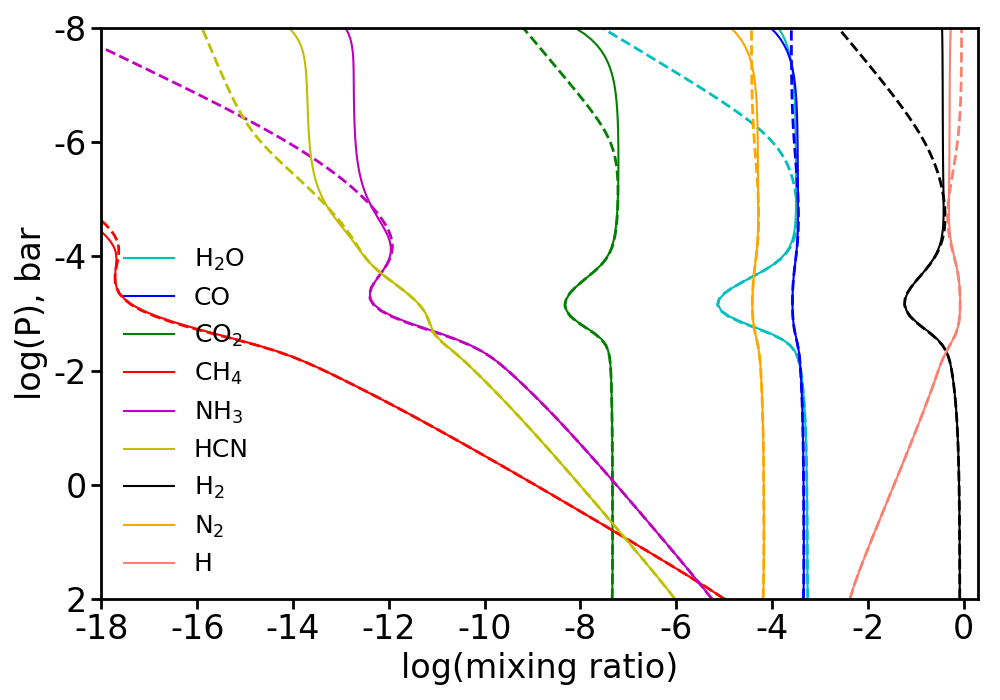}
\includegraphics[width=\hsize]{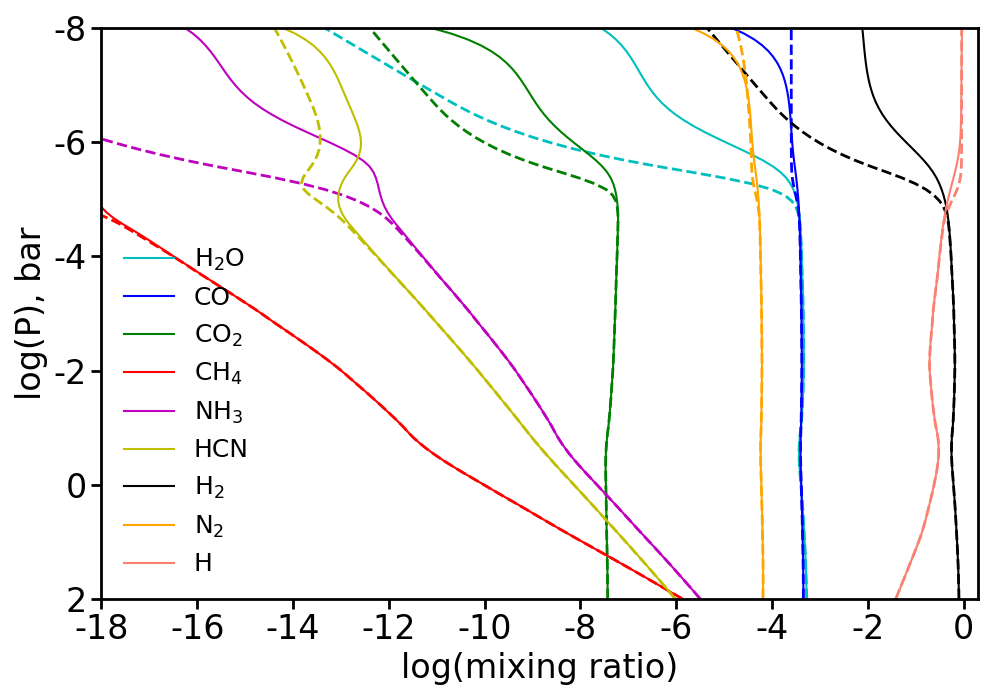}
\includegraphics[width=\hsize]{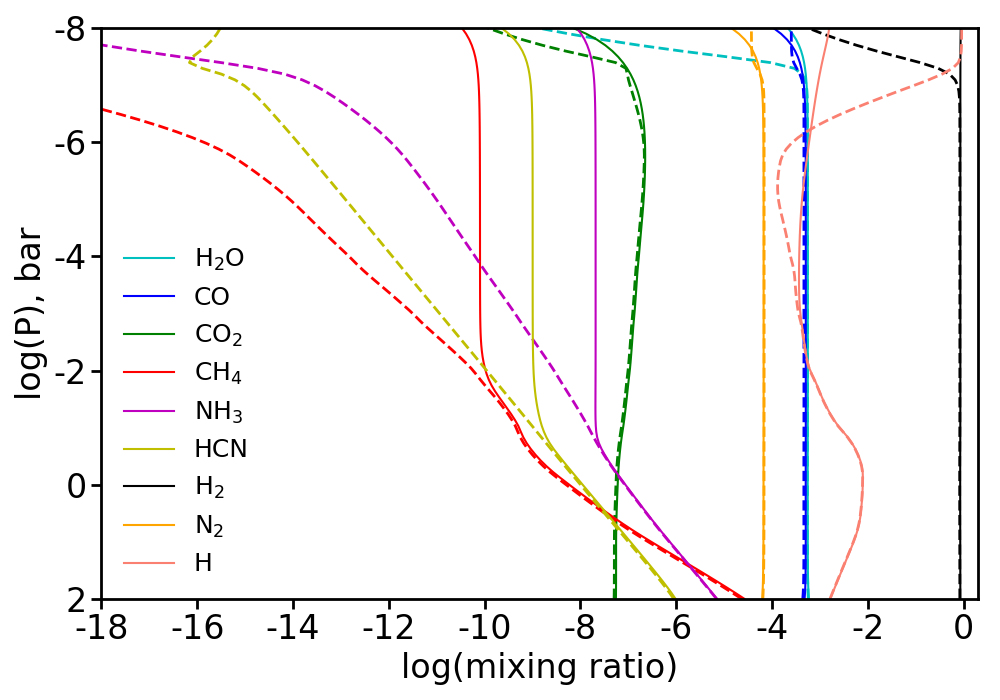}
\includegraphics[width=\hsize]{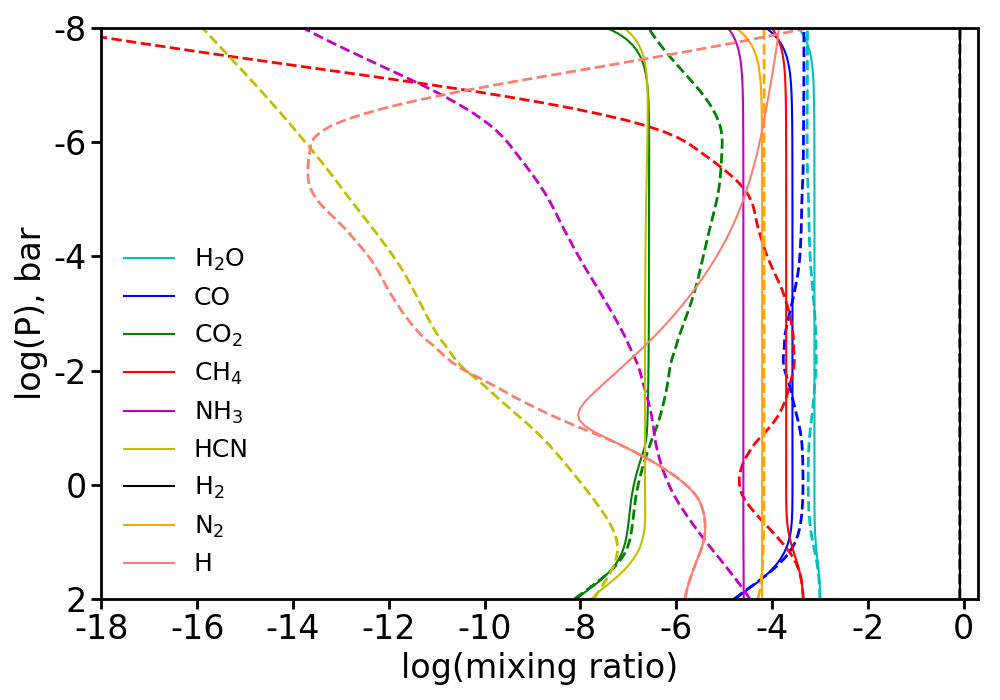}
\end{minipage}
\caption{Same as on Fig.~\ref{fig:sp-vmr-a0}, but for the planet around a K0 star.}
\label{fig:sp-vmr-k0}
\end{figure*}

\subsection{Atmospheric chemistry}
We first look at the influence of stellar spectral types on the chemical composition.
The corresponding plots are shown on the right panels of Figs.~\ref{fig:sp-vmr-a0}, \ref{fig:sp-vmr-f0}, \ref{fig:sp-vmr-g2}, and~\ref{fig:sp-vmr-k0}.
First, we find that, regardless of the star type,  for the coolest planet
considered here, i.e. T1 case, the disequilibrium {processes}
can affect the mixing ratio of molecules \nhhh, \chhhh, and \hcn\ throughout
the whole atmosphere and even down to optically thick layers which lay below 1~bar {level}. 
Also \citet{2011ApJ...737...15M} noted that for HJs HD~209458b and HD~189733b
(planets orbiting main-sequence G0 and K2 stars, respectively, with 
$\teq$$\sim$1200~K), molecules as \coo, \co, \hho, and \nn\ are relatively 
unaffected by disequilibrium chemistry. Our study extends this conclusion 
to hotter stars of spectral types A0 and F2.

When the planet is placed closer to the host star (i.e., cases T2 and T3), 
the higher XUV radiation by the A0 star compared to other stars
considered in this model (see Fig.~\ref{fig:stellar_fluxes}) induces a more efficient molecular 
photodissociation. This also produces a strong decrease of the 
species mixing ratios at low pressure levels where the radiation 
can penetrate. The characteristic sudden drop
in molecular mixing ratios seen at around P$\approx 10^{-3}$~bar for the
T3-inverted case (top right plots on Figs.~\ref{fig:sp-vmr-a0}-\ref{fig:sp-vmr-k0}) is caused by the
corresponding temperature inversion that is present in the parameterized 
temperature profile (see Fig.~\ref{fig:t1t2t3}). Here, the mixing
ratios of all species are efficiently quenched in and below the region 
of temperature inversion. Therefore, the decrease
in mixing ratios is dictated by thermochemistry in this hot
environment at the region of thermal inversion. For all DQ models and temperatures explored
in this work we find that \co\ stays the dominant carbon bearing 
molecule as it is efficiently formed by thermochemical 
processes and barely affected by either chemistry or photodissociation. 
However, for the T1 case the concentration of \chhhh\ becomes comparable to
that of CO for the HJ orbiting cool G2 and K0 stars. This is because 
a) thermochemical equilibrium renders similar mixing ratio values at 
those temperatures, and b) the XUV flux emitted from these stars is 
very weak in our photospheric models to dissociate these species
(see bottom right plots on Fig.~\ref{fig:sp-vmr-g2} and Fig.~\ref{fig:sp-vmr-k0}). 
Strong UV flux is
needed to noticeably distort the concentration of \nn\ away from
its thermochemical equilibrium value, and it stays almost unchanged in mid and
low altitudes for all spectral types except A0. Molecular hydrogen 
remains the most abundant species (except at very high atmospheric layers) 
for temperatures $\teq$$<$3000~K and in all spectral
types. As temperatures increase (i.e., as the exoplanet orbits closer to the host star),
concentration of neutral hydrogen 
raises rapidly due to an efficient thermal dissociation 
and photodissociation of hydrogen bearing molecules. But also the concentration of \hcn\ raises for
T1 planet compared to the EQ calculations for every star considered
in this work. This is in full agreement with previous studies
by \citet{2011ApJ...737...15M} and  \citet{2019MNRAS.487.2242H} for HD~209458b and 
HD~189733b, although their analysis only pertained to G and K stars
and ours pertains to a broader range of star types.  
When the photochemistry is included in the calculations, it largely 
influences the distribution of species in optically thin layers where 
the deposit of stellar XUV radiation is very efficient (and this effect is individual for 
each molecule). The two commonly used diffusion coefficients 
in 1D photochemical models (i.e., eddy and molecular diffusion coefficients)
can determine the variation of the mixing ratio profiles with pressure if
the characteristic diffusion times are shorter than the chemical lifetimes.
Otherwise, the vertical distribution of species will be mainly 
controlled by photo/thermochemical processes. Note that there might exist
regions where the associated lifetimes are similar and they can compete for
shaping the mixing ratio profile.
However, a detailed analysis of the prevalence of chemical and/or transport processes
as a function of the star type, of the choice of the Kzz, and of temperature of the 
exoplanetary atmosphere is beyond the scope of the current paper.

Finally, our calculations agree with previously reported conclusion that the impact of disequilibrium chemistry
is stronger for cooler planets due to the higher concentrations of photochemically active gases like \chhhh\ and \nhhh~\citep{2016SSRv..205..285M,2014RSPTA.37230073M}.

\subsection{Predicted transmission and emission spectra}

The left column of Figs.~\ref{fig:sp-vmr-a0}, \ref{fig:sp-vmr-f0}, \ref{fig:sp-vmr-g2}, and~\ref{fig:sp-vmr-k0} shows
transmission and emission spectra calculated using EQ and DQ mixing ratios discussed above.

In our simulations the effect of spectral type is to provide non-thermal XUV radiation which
sets the rates of molecular {photo-dissociation
while the temperature has effects on the transport processes and on the T-dependent chemical reactions.}
Hence, we detect most prominent changes between EQ and DQ models in atmospheric spectra 
for the hottest stellar type A0 and planets with T1 and T3-inverted temperature profiles
(Fig.~\ref{fig:sp-vmr-a0}, where we highlighted spectral features of various chemical species 
across a range of wavelengths that are useful for identifying their signatures for the T1 case in transmission).
For the T1 case, the DQ model predicts lower flux in the region between 10~\mum\ and 20~\mum\ in the emission spectrum. 
This is the result of the increased \nhhh\ concentrations
and therefore stronger absorption in these wavelengths compared to EQ model. Absorption features 
at 10.4~\mum\ and 14~\mum\ that are seen in DQ models are due to \nhhh\ and \hcn\ lines, respectively.
The increase of the spectral emission in DQ model compared to EQ one 
between 3~\mum\ and 4~\mum, and between 7~\mum\ and 9~\mum\ is due to photodissociation
of \chhhh\ and hence weakening of the line absorption. 
In transmission the planet look larger in the wavelengths of increased opacity, i.e.
in the region between 10~\mum\ and 20~\mum\ due to the increased \nhhh\ concentrations predicted by DQ model.
At the wavelength shorter than that the transmission signal
is decreased primarily due to a weaker absorption by \chhhh. A pronounced feature at 4.3~\mum\ is due
to enhanced strengths of \co\ and \coo\ bands. An increase in the transmission signal
around 1.6~\mum\ and 2.1~\mum\ is due to the increased absorption in \nhhh\ bands located at these regions.
We detect same spectral behavior for the T1 planet around F0, G2, and K0 stars.
{Note that many expected features have been already identified in previous investigations
\citep[e.g.,][]{2011ApJ...737...15M,2012A&A...546A..43V,2018ApJ...853..138B,2019ApJ...883..194M}, 
and our study extends this analysis to a wider range of stellar spectral classes and atmospheric temperatures.}

{For the T2 planet both emission and transmission spectra show almost no changes between EQ and DQ models for all spectral types.}
{In case of T3 planet we detect a noticeable decrease of the transmission amplitude but only for A0 star which is mainly due
to a strong photo-dissociation of \hho\ (but not only).}
This is an interesting finding because it indicates that for planets
with $\teq$$\approx$2000~K strong differences in mixing ratios that are driven by the radiation of the host star could not be robustly detected
in the observed spectra. In case of emission radiation, the photodissociation affects mixing ratios of atmospheric species
at high altitudes that are transparent for the outgoing radiation and therefore do not significantly contribute to the emerging spectrum. 
Also, the abundance of \chhhh, \nhhh, and \coo\ is very low even in EQ model so that the changes in their
mixing ratios introduced by photo-dissociation are not manifesting in the final spectra. Strong photo-dissociation of \hh\ results
in two orders of magnitude increase in the number density of neutral hydrogen atoms, but the corresponding opacity 
(due to Rayleigh scattering on H atoms and \hminus\ bound-free and free-free absorption) 
is not strong enough to drive changes in the predicted spectra either.

For the more realistic case of an UHJ with an inverted temperature profile (T3-inverted model) we observe strong differences
between EQ and DQ models which is driven by molecular photo-dissociation, but again only for A0 star. 
{As a result, the amplitude of spectroscopic features due to, e.g., \hho\ and \co\ are strongly reduced
in transmission. In emission, the DQ model emits more flux thus making the planet look brighter.} 
Note that the short wavelength
part of the spectrum is not affected by disequilibrium chemistry and thus EQ and DQ models look the same
for $\lambda<1$~\mum\ in case of transmission and $\lambda<2$~\mum\ in case of emission, respectively.

For the planets around F0, G2, and K0 the effect of disequilibrium chemistry is strongly reduced for T2, T3, and T3-inverted
cases. This is because of a weakening of the stellar XUV flux compared to A0 star, as well as very low mixing ratios of spectroscopically
active molecules such as, e.g., \chhhh, \nhhh, and \hcn\ predicted by EQ and DQ models, {in agreement with
the predictions from previous studies \citep[e.g.,][]{2016SSRv..205..285M}.}

\subsection{Impact of stellar activity}

\begin{figure*}[!ht]
\centerline{
\includegraphics[width=0.49\hsize]{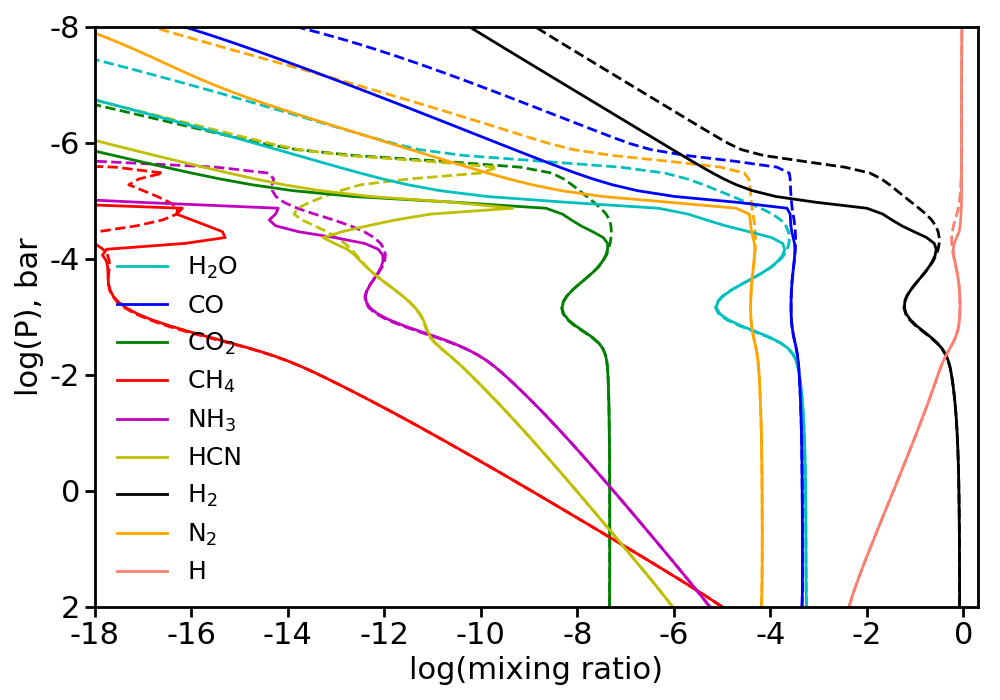}
\includegraphics[width=0.49\hsize]{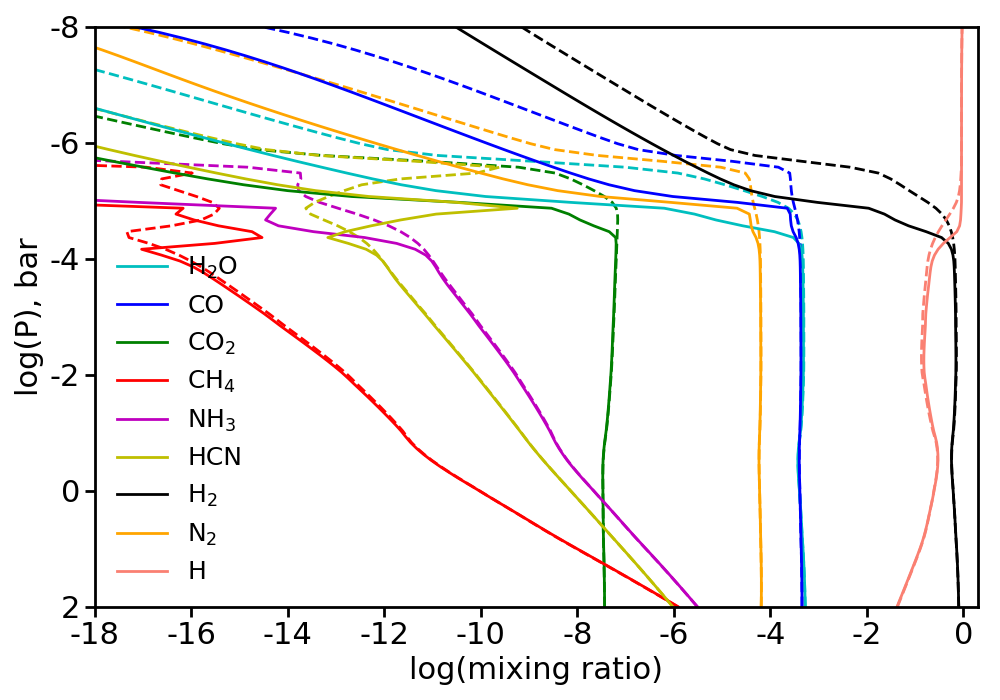}
}
\centerline{
\includegraphics[width=0.49\hsize]{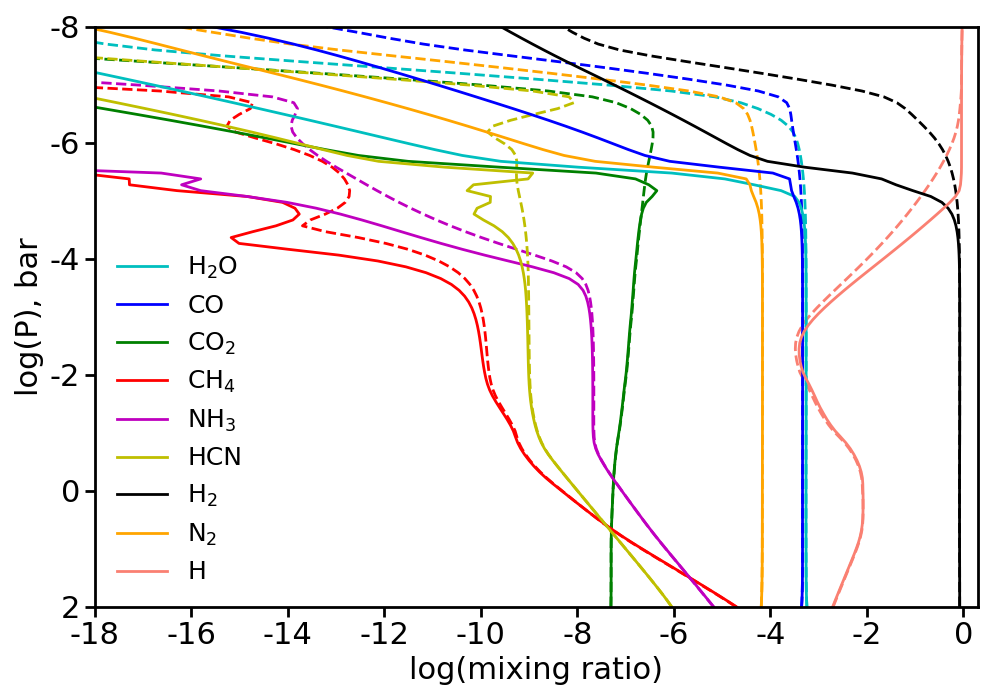}
\includegraphics[width=0.49\hsize]{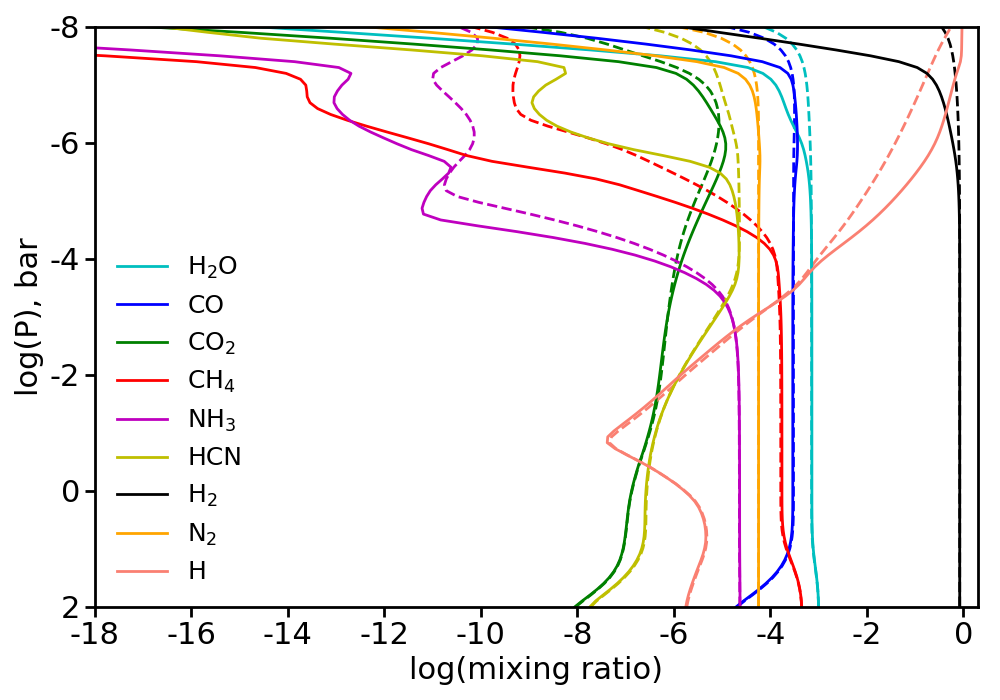}
}
\caption{Volume mixing ratios in the atmosphere of a Jupiter size planet around present (dashed lines) and 0.1~Gyr young Sun (solid lines)
for the planet with equilibrium temperatures of $\teq$$=$3000~K with temperature inversion (top left), $\teq$$=$3000~K (top right),
$\teq$$=$2000~K (bottom left), and $\teq$$=$1000~K (bottom right).}
\label{fig:sp-vmr-youngsun}
\end{figure*}

To this end we used synthetic flux predicted by models of stellar photospheres to compute disequilibrium mixing ratios 
with \vulcan\ ignoring the activity driven enhancement of XUV radiation.
As we have noted before stars with convective envelopes generate magnetically driven activity that manifests itself
as an enhanced XUV radiation due to the non-thermal atmospheric heating.
Stars also undergo changes in their activity level as they age.
Thus, in this section we investigate the impact of these two effects on the chemical structure of the atmosphere
of our test planet.

\begin{figure*}
\begin{minipage}{0.59\hsize}
\centerline{
\includegraphics[width=\hsize]{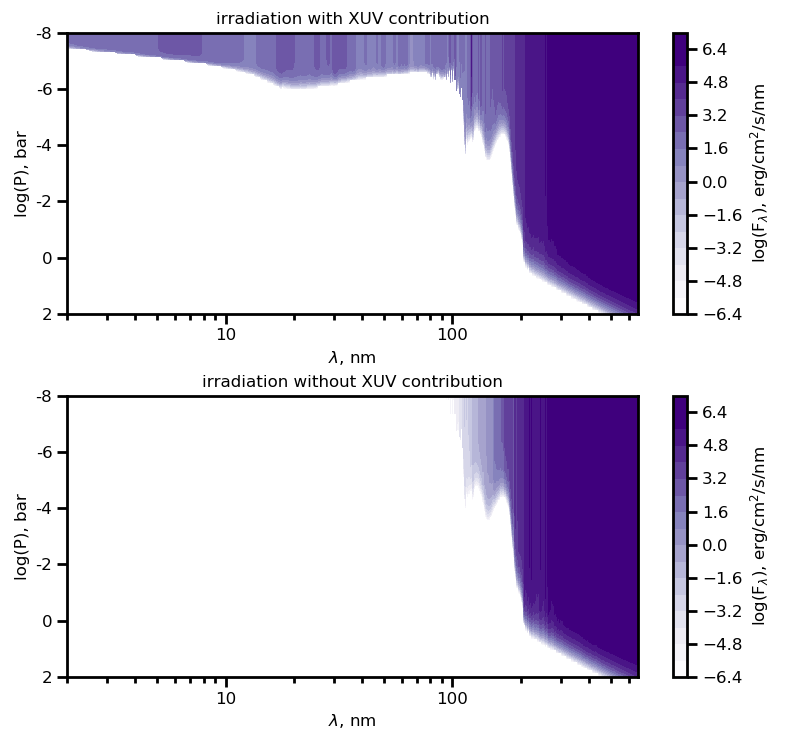}
}
\end{minipage}
\begin{minipage}{0.40\hsize}
\includegraphics[width=\hsize]{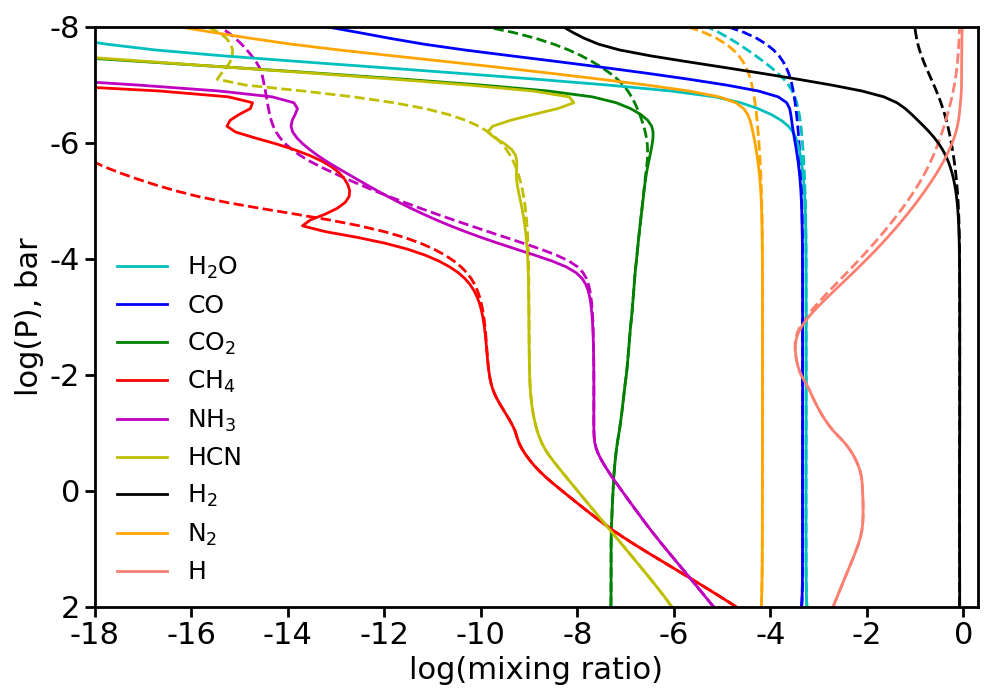}
\end{minipage}
\caption{Absorption of stellar flux in the atmosphere of a planet with $\teq$$=$2000~K as a function of wavelength
(left panel) calculated using observed solar radiation (top panel) and stellar flux without activity driven XUV contribution 
as predicted by photospheric \phoenix\ model (bottom panel). The corresponding mixing ratios are shown on the right panel
(dashed line~--~using \phoenix\ flux, solid line~--~using observed solar flux).}
\label{fig:sp-vmr-g2-sunmodern}
\end{figure*}

First we compared predictions for the present Sun with its measured activity level 
and calculations for the G2 star without activity driven XUV contribution by using \phoenix\ photospheric flux.
As an example, Fig.~\ref{fig:sp-vmr-g2-sunmodern} illustrates the amount of stellar flux
that reaches different atmospheric depths as a function of wavelength for the T2 case.
It is seen that most of the XUV flux of the modern Sun is efficiently absorbed at very high altitudes.
In this particular T2 case the radiation
with wavelengths longer than about 200~nm is capable of reaching very deep atmospheric layers,
while flux at shorter wavelengths is fully absorbed at layers above $10^{-6}$~bar. As a result,
we detect the strongest changes in species mixing ratios (compared to equilibrium calculation) 
in these high altitudes, as shown on the right panel of Fig.~\ref{fig:sp-vmr-g2-sunmodern}
(For the hottest T3 planet we find the same distribution of the absorbed radiation
but the XUV radiation reaches $\sim10^{-5}$~bar).
Because mixing ratios remain unaffected by stellar radiation at mid- and deep atmospheric layers,
we detect almost no changes in the predicted transmission and emission spectra.
We thus conclude that one must always account for the stellar activity contribution to the XUV flux in order to capture important changes 
in species concentrations of planets orbiting stars with convective envelopes. Strong photo-dissociation at these high altitudes
increases concentrations of light species and thus enhance atmospheric erosion \citep[e.g.,]{2019A&A...624L..10J}.
However, normal photospheric stellar models can still be used to predict planetary atmosphere spectra, 
at least in cases when the parent star is not at its strongest activity state.

Next, we analyzed the impact of enhanced stellar activity
on our test planet assuming that it orbits the present Sun with its measured XUV flux
and around the Sun when it was 100~Myr young. At that young age the XUV flux of the Sun was maximal
thus driving strongest changes in the atmospheric chemistry which we compare on Fig.~\ref{fig:sp-vmr-youngsun}.
We find that when activity is included, the \co\ still remains the dominant carbon bearing molecule for all planetary temperatures considered
{due to low photodissociation efficiency}. 
That is, even very high XUV and $\lalpha$ radiation are not able to photodissociate \co\ in lower and middle atmospheric layers.
Only at high altitudes there is enough stellar flux that finally triggers dissociation of \co.
For the T1 case, the number
density of \chhhh\ also becomes relatively high in altitudes up to about 10$^{-4}$ bar. 
Generally, for all planets we observe strong photodissociation of molecules at high altitudes of the planet caused
by increased stellar activity at dissociating wavelength ranges. This effect is more noticeable for planets closer to the host
star, i.e., T2, T3 and T3-inverted cases.
For the T3 and T3-inverted cases molecules \chhhh\, and \hcn\, show {noticeably variable profiles} at altitudes above $10^{-4}$~bar, with even increased
concentrations in narrow pressure region.

Despite large changes in molecular mixing ratios between modern and young Sun calculations, we find no particular features 
in the transmission and emission spectra that could be used to estimate the age and therefore activity state of the host star (not shown here).
Only the amplitude of the spectra is modified due to the fact that the young Sun was 30\%\ dimmer and 12\%\ smaller compared to its present values.

\subsection{Constraining disequilibrium {processes} with future missions}

\begin{figure*}[!ht]
\centerline{
\includegraphics[width=0.5\hsize]{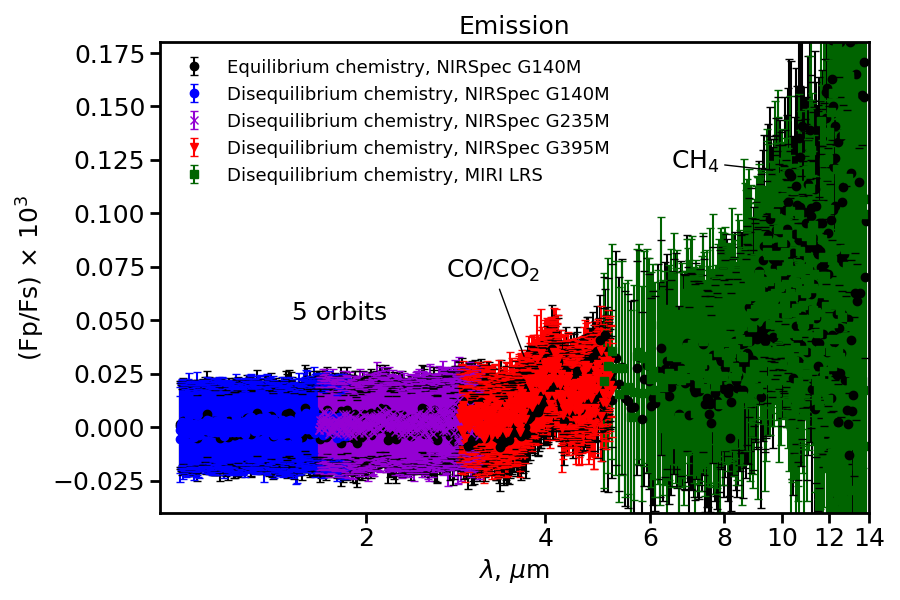}
\includegraphics[width=0.5\hsize]{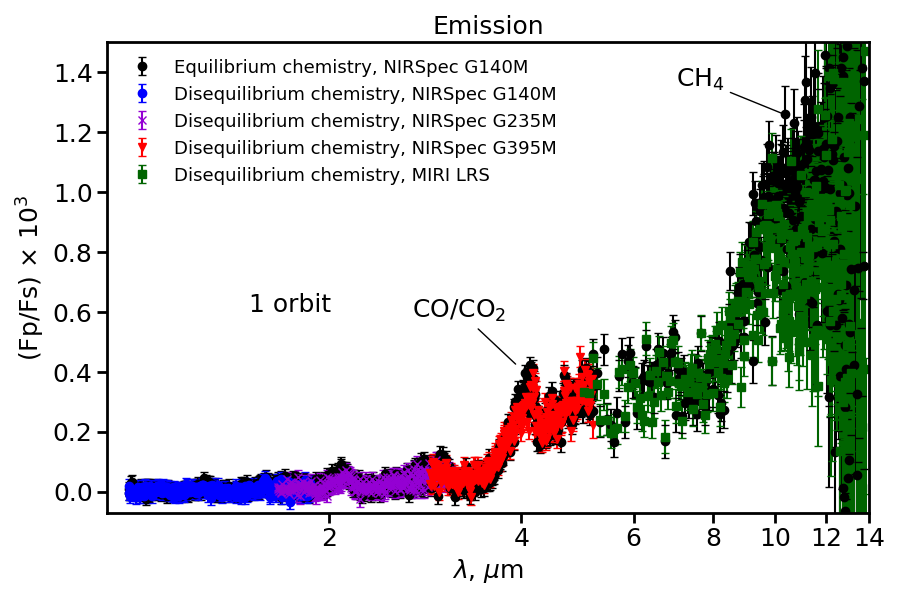}
}
\centerline{
\includegraphics[width=0.5\hsize]{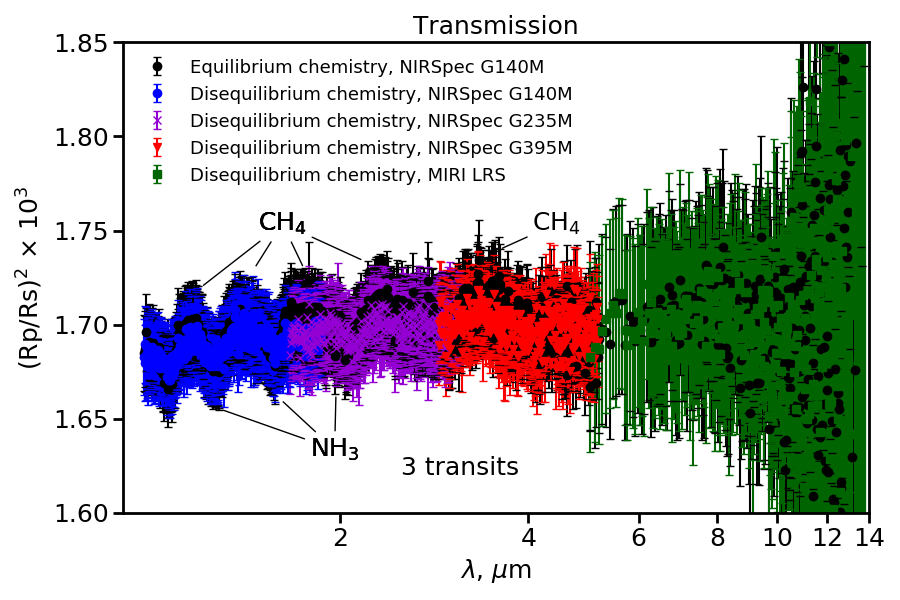}
\includegraphics[width=0.5\hsize]{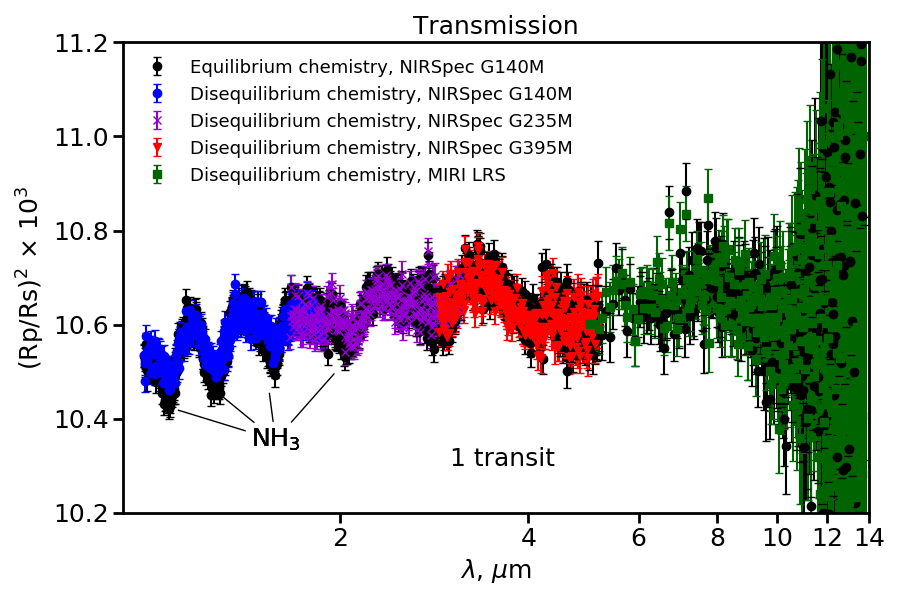}
}
\caption{Simulated emission (top row) and transmission (bottom row) spectra for a hot Jupiter planet with $\teq=1000$~K orbiting A0 (left) and G2 (right) stars
using NIRSpec and MIRI instruments with JWST. 
The molecular features produced by photo-kinetic models that could be unambiguously detected with corresponding
number of transits are explicitly marked.}
\label{fig:jwst-a0-sun-vpl}
\end{figure*}

\begin{figure*}
\centerline{
\includegraphics[width=0.5\hsize]{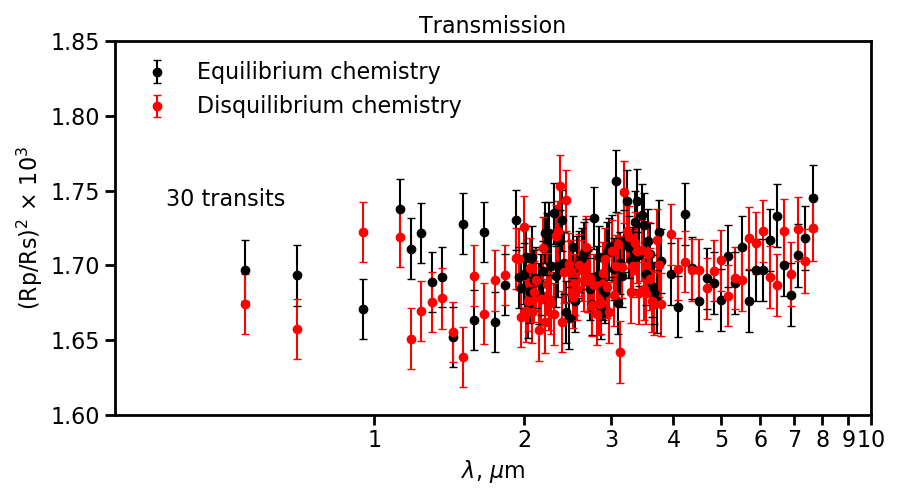}
\includegraphics[width=0.5\hsize]{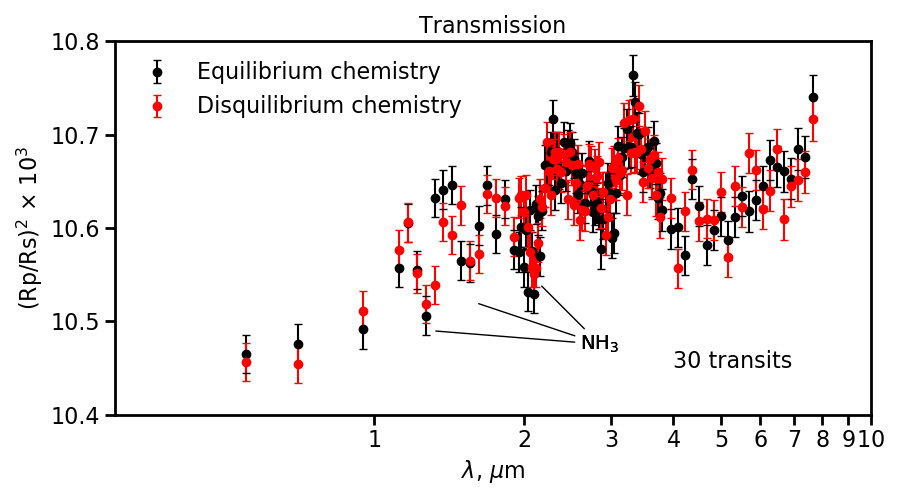}
}
\caption{Simulated ARIEL transmission spectra of a $\teq=1000$~K planet orbiting A0 (left) and G2 (right) star.}
\label{fig:ariel-trans}
\end{figure*}

As was shown above, the DQ models for HJs with $\teq=1000$~K produce characteristic strong features in transmission and emission spectra
that clearly distinguish them from EQ models. These features could be studied with future space missions and the detection 
or, alternatively, non-detection of these features can provide strong constrains for modern photo-kinetic models, but not only.

To provide estimates of the observability of photochemically produced spectral features 
we used \textsc{PandExo}\footnote{\tt https://natashabatalha.github.io/PandExo}
data simulation platform \citep{2017PASP..129f4501B} and carried out predictions 
for the James Webb Space Telescope (JWST)\footnote{\tt https://www.jwst.nasa.gov}.
Figure~\ref{fig:jwst-a0-sun-vpl} shows predicted emission and transmission spectra of our test planet
around A0 and G2 stars for the T1 case. We chose these two {stellar} types because this is where 
the impact of disequilibrium chemistry is the strongest. We made predictions for the NIRSpec instrument with 
medium resolution mode (gratings G140M, G235M, and G395M) and MIRI instrument in low resolution mode (LRS). 
The large noise at wavelengths $\lambda$$>$10~\mum\ is due to a dramatic drop in the MIRI-LRS throughput (Natasha Batalha, priv. comm.).
The realistic noise floor for JWST instruments will be known not until after the instrument is operational,
and similar to previous studies \citep{2017PASP..129f4501B,2016ApJ...817...17G} we assumed it to be
20~ppm and 50~ppm for the NIRSpec and MIRI-LRS modes, respectively. This noise floor was added as a quadrature
to the photon noise generated by \textsc{PandExo} to obtain final errors on simulated data.
We further binned the spectra to the resolution of $R=\lambda/\Delta\lambda$=300
to boost S/N in each channel (excluding noise floor). 
The stars were assumed to have K-band magnitude 5$^m$ ($\lambda_{\rm ref}=2.22$~\mum).
The transit duration was 4~h and 15~h for the G2 and A0 cases, with planet orbital periods being 8~d and 114~d, respectively.
The total integration time for a single transit/orbit (without overheads) was
8~h and 12~h for G2 stars and 30~h and 45~h for A0 star in transmission and emission, respectively.
We then adjusted the number of corresponding orbits/transits needed to reach a desired spectroscopic accuracy to robustly
detect features caused by disequilibrium processes.

We found out that the noise floor is easily reached in many wavelength channels already in few orbits/transits
(especially in G140M and G235M wavelengths, and to a lesser extent in G395M),
and thus the binning of the data to even lower resolution does not help to decrease final errors.
The noise floor is never reached in MIRI-LRS within the integration time considered in our simulations, 
and the intrinsic instrument performance thus sets the limit on the final errors.
However, the detection of features due to disequilibrium processes would still be possible thanks to a large number
of wavelengths that can be observed simultaneously. We estimate the detectability of spectroscopic features
by using $\chi^2$ test between original spectra of EQ and DQ models and simulated DQ observations, respectively. 
The corresponding values of $\chi^2$ were computed only within spectral 
intervals affected by disequilibrium processes.

Our estimates show that it would be challenging to spectroscopically observe photochemical signatures 
in the atmosphere of a planet orbiting A0 star because the star-to-planet
size ratio as well as the flux contrast is very unfavorable at near-infrared wavelengths. 
In emission observations, five orbits would be needed
to detect the effect of disequilibrium chemistry using \co/\coo\ feature at 4.3~\mum\ above 3$\sigma$ threshold. 
In transmission, many disequilibrium effects
could be seen with 3 transits, including weak \nhhh\ features (see annotations on Figure~\ref{fig:jwst-a0-sun-vpl}).
On the other hand, the decrease in the transmission amplitude at 1.8~\mum\ and 2.23~\mum\ due to the dissociation of \chhhh\ could already be 
detected with even fewer number of orbits. Recall, however, that obtaining transmission spectrum even for this very bright A0 star
takes 30~h of integration time which is split between 15~h in- and 15~h out-of-transit observations.

{For the G2 star case, the robust detection of the  \co/\coo\ feature  at 4.3~\mum\ and \chhhh\ at 11~\mum\ 
above 3$\sigma$ level requires only one orbit in emission observations, as shown on the top right plot on Fig.~\ref{fig:jwst-a0-sun-vpl}.
In transmission, one transit is needed to detect \nhhh\ features, but detecting the \co/\coo\ feature
at 4.3~\mum\ and \nhhh\ absorption at 10~\mum\ remain challenging and will require larger number of transits for robust analysis.}

The \hcn\ feature seen at 14~\mum\ for A0 case can potentially be detected with the MIRI@JWST instrument 
(channel 3, wavelength rage 11.53-18.03~\mum) using the medium resolution integral field unit mode (MIRI MRS).
Because the current version of the \textsc{PandExo} does not support this observing mode, we could not
calculated expected noise level and thus required observing time. However we notice that amplitude of the \hcn\ band
is relatively strong compared to other photochemically produced features and thus could be addressed with JWST.

For the T3-inverted cases of an A0 star, although the difference between EQ and DQ models is large, it would be fairly difficult
to study photochemical processes because the predicted changes in the transmission spectra go the same way in a wide wavelength range.
That is, the DQ model looks just like EQ one with decreased transmission amplitude (see Fig.~\ref{fig:sp-vmr-a0}). 
Nevertheless, with accurate enough observations this decrease in the amplitude of spectral features could also be studied by comparing
flux at short and long wavelengths and/or by analyzing the amplitude of the molecular bands between 3-4~\mum\ and 6-7~\mum, respectively.
{From simulations listed above we conclude that the spectroscopic accuracy of JWST will be enough
to study strong changes in the transmission amplitude in a wide wavelength range cased by disequilibrium processes
in case of inverted temperature profile (T3-inverted case).
However, more detailed calculations are still needed to compute electron number densities and thus \hminus\ opacity self-consistently.}
Overall we find that NIRSpec@JWST will be just the right instrument to study individual details in spectra of extrasolar planets due to its
sufficient spectral resolution and wide wavelength coverage.

Low resolution spectroscopy will also be available at the Atmospheric Remote-sensing Infrared Exoplanet Large-survey
space mission (ARIEL)\footnote{\tt https://arielmission.space/} planned by ESA for 2028 \citep{Tinetti2018}.
The instrument will have few photometric filters between 0.5~\mum\ and 1~\mum, and two spectroscopic options
with resolving power R$=$100 between 1.95-3.95~\mum\ and R$=$30 between 3.95-7.8~\mum\ thus covering a spectral region
where the signatures of disequilibrium chemistry can be studied.
{We used the latest ARIEL noise simulator (L.~Mungai and E.~Pascale, priv. comm.) to estimate the noise level
in instrument pass bands.
As an example, Fig.~\ref{fig:ariel-trans} shows the transmission spectra (with random noise added) 
of our test planet around A0 and G2 stars assuming the same transit duration and stellar magnitude as above for the JWST.
Unfortunately, a relatively small number of available wavelengths and a small size of the telescope
makes it virtually impossible to constrain disequilibrium processes
because the 3$\sigma$ detection threshold is almost never reached. For A0 star, even after 30 transits the DQ model
deviates only by 1$\sigma$ from the EQ one with the noise floor dominated.
With the same 30 transits we do reach on average 3$\sigma$ threshold for the G2 star, but more transits
would not help due to hitting the noise floor. Note that studying disequilibrium processes might still be possible
for planets orbiting low-mass stars.}

\section{Discussion}

\subsection{Consistency between atmospheric temperature and mixing ratios}
In our calculations of the impact of disequilibrium chemistry on the observed emission and transmission spectra
of HJs we used the P-T structure generated with \helios\ but assuming chemical equilibrium concentrations. The latter
were calculated with the \fastchem\ and were used to compute continuum and line opacity needed for \helios.
However, as initial equilibrium mixing ratios are changed by photo-chemistry and kinetics, they also change
local opacity and therefore local temperature which, in turn, changes mixing ratios. 
This means that in our approach the planetary T-P structure is inconsistent with
the results of disequilibrium chemistry calculations. For instance, \citet{2016A&A...594A..69D} 
found that for such well known planets as HD~189733b ($\teq=1200$~K) and HD~209458b ($\teq=1500$~K)
the effect of non-including consistent calculations between chemistry and temperature  
leads to an overestimation of the impact of disequilibrium chemistry on the predicted emission spectra.
This happens because once temperature iteration is introduced in the disequilibrium chemistry algorithm, the model
P-T structure adapts to the evolution of mixing ratios in order to maintain the global energy balance in the planetary
atmosphere (which is set by the corresponding $\teq$ of the planet). Therefore, eventhough the final mixing ratio profiles
and P-T distribution look different from their initial equilibrium states, the predicted emission spectra does not deviate very much from
the equilibrium calculations. It does however contain some distinct features and thus disequilibrium  chemistry must
be taken into account for accurate interpretations. For instance, the \nhhh\ feature at 10~\mum\ is present in both
purely photochemical and self-consistent photochemical models \citep[see Fig.~7 in][]{2016A&A...594A..69D}. This feature
is also prominent in our simulations of the emission spectra of the T1 planet and thus should be robustly detected
in future observations.

\subsection{Atmospheric opacity}
In our calculations of \hminus\ opacity we used the equations given in \citet{1988A&A...193..189J}
for bound-free and free-free processes. These equations require the knowledge
of \electron\ and \ion{H}{i} number densities. Unfortunately, at its current stage
\vulcan\ does not account for the ionization of atomic species. Therefore, we used
the concentrations of electrons calculated by \fastchem\ and concentrations of \ion{H}{i} from \vulcan\ to compute \hminus\ opacity.
We do not expect this inconsistency to affect our results much. At least it should not be an issue
for T1 and T2 cases where the atmospheric temperatures are too low 
for the \hminus\ and \heminus\ opacity to play any significant role. However, for T3 and T3-inverted cases
a more accurate calculation of \electron\ concentrations is desired, especially in dense and hot layers
around photosphere and below.

\subsection{Stellar activity: XUV radiation}
In this work we explicitly addressed the impact of stellar activity on the planetary spectra of a G2 star using observed XUV flux of the modern Sun as a proxy.
We then additionally investigated the changes in mixing ratios and spectra of a young Sun of age 0.1~Gyr. 
This is the youngest age available from the study by \citet{2012ApJ...757...95C} 
and at that age the Sun had the highest XUV flux so that the changes in photochemistry were the strongest. 
Note that, according to \citet{2015A&A...577L...3T}, sun-like stars with ages between 20~Myr and 300~Myr maintain very high X-ray flux which is almost
constant with time and does not depend on stellar rotation rate and hence the age.
After 0.3~Gyr the activity in a majority of sun-like stars begin to dissaturate and their X-ray flux 
starts to decay with time as rotation rate keeps decreasing due to the magnetic braking.
Thus, in principle we could have chosen any parameters for the young Sun within this time interval
but this would not have changed the general trends in our results.

One important caution needs still to be made when addressing activity in young suns. 
Observations of stellar clusters demonstrated that sun-like stars are born with a rather wide range of rotation periods \citep[e.g.,][]{1993ApJ...409..624S}
and it is not until about 1~Gyr that they all presumably converge to the same rotation period of about 10 days \citep{2015A&A...577L...3T}.
Because the age at which the stellar activity stars to dissaturate depends on the initial rotation period, 
sun-like stars that are younger than 1~Gyr could still have very different XUV flux
depending whether they were born as fast or slow rotators \citep{2015A&A...577L...3T}. 
That means that, e.g., for the Sun it is currently impossible
to predict the true value of its XUV flux when it was young because we do not know how fast the young Sun was rotating
in a first place. However, this is not an issue in our study because we analyzed a hypothetical gas giant planet around a young sun-like star
and we assumed that this star was very active in its early evolution stages with best-to-date estimates of the expected XUV radiation
taken from \cite{2012ApJ...757...95C}. It is obvious that studying chemical evolution of atmospheres of real planets 
must account for the accurate knowledge of the activity evolution of their hosts.

In this study we did not consider the impact of activity for the planets orbiting K0 star. Indeed, as was mentioned above,
all stars with convective envelopes generate non-thermal XUV emission similar to our Sun. What is different from sun-like stars though 
is the less efficient magnetic braking of K and M stars because of their small sizes \citep{2012ApJ...746...43R}. 
As a result, K and M stars can maintain strong 
activity level for a much longer time compared to sun-like stars with a potentially stronger long-term impact on atmospheric chemistry.
Moreover, M dwarfs are known to generate strongest magnetic fields among all stars with convective envelopes \citep{2017NatAs...1E.184S}
resulting in extreme magnetic heating of their chromospheres and coronae. For instance, the ratio of X-ray to total luminosity of M dwarfs 
are 10 to 100 times higher than those of the present day Sun \citep{2010A&A...515A..98P} whereas their bolometric
luminosity are 10 to 1000 times smaller. Reconstructing XUV spectrum of these stars is a difficult task. Nevertheless, for a dozen of planet host 
targets the combined X-ray, UV, visual, and infrared data exist and available from, e.g., the \textsc{MUSCLES}
survey\footnote{\tt https://archive.stsci.edu/prepds/muscles/} \citep{2016ApJ...820...89F,2016ApJ...824..101Y,2016ApJ...824..102L}. 
These data could be use to compute realistic photochemical models of planets
orbiting these small stars. However, we expect that the activity of K stars affects the observed spectra in a similar way it did for the case of a young Sun, though
the amplitude of the effect could be stronger due to an increase in a relative X-ray luminosity and the fact the our test planet would be closer to the star
(to maintain equilibrium temperatures that we assumed in this work). 

As stars spin down their chromospheric activity eventually fades away. But even in this case old stars with convective envelopes
can still have non-zero X-ray emission, this time of non-magnetic nature, which is called \textit{basal emission} \citep{1995A&ARv...6..181S,1989ApJ...341.1035S}
and which can be a major source of high energy photons that drive planetary atmospheres out of equilibrium even if stellar dynamo is very weak or even absent.

\subsection{Stellar activity: winds}
Winds are another aspect of stellar activity that may have a strong impact on the atmospheric structure at high altitudes.
Stellar winds carry high energetic particles that collide with atmospheric constitutes, dissociate and ionize them,
and thus raise local atmospheric temperatures. Unless planets have strong magnetic fields that protect their atmospheres against
stellar winds, the latter cannot be ignored, especially for active stars. Building realistic wind models
for different type of stars and ages is {very challenging because} many unknown parameters that often rely on very limited observations
\citep[see][and references therein]{2014MNRAS.443..898L}. Nevertheless, some sophisticated wind models have been developed that account
for outflows dominated by coronal mass ejections (CMEs) \citep{2015A&A...577A..27J,2015A&A...577A..28J} and/or 3D MHD wind models
that can take into account the observed properties of stellar magnetic fields \citep{2015MNRAS.449.4117V}.

Overall, it is seen that the impact of stellar activity on the atmospheric structure of planets must include many
physical {and chemical} processes that are not trivial to model and often even estimate. In this work we limited ourselves to only stellar XUV radiation. 
Addressing activity impact in real stars would surely require consideration of all these processes
to the best of our knowledge.

\subsection{Clouds and hazes}
In this work we ignored the possible contribution of clouds and hazes to the observed planetary spectra.
It was shown that some HJs may have thick cloud decks at high
enough altitudes that are probed with transmission spectroscopy \citep{2016Natur.529...59S}.
The effect of clouds would weaken spectroscopic features due to scattering absorption.
This effect is the largest at optical wavelengths and decays for longer ones.
Thus, the presence of clouds and hazes can mute the photochemically produced features of \nhhh\ between 1~\mum\ and 3~\mum.
Nevertheless, clouds are not expected to seriously affect spectra of planetary atmospheres at $\lambda>10$~\mum\ 
(except perhaps condensate vibrational mode features which can become prominent in the infrared \citep{2015A&A...573A.122W}),
thus favoring analysis of disequilibrium effects in that spectral domain.

\subsection{{3D atmospheric effects}}
{
In this work, we have considered a 1D photochemical model representative of mid-latitudes and globally-averaged diurnal conditions. 
Also, the same temperature structure for the day and night sides of our test planet were assumed when predicting its spectra. 
In reality, however, chemistry in atmospheres of HJs could be strongly modified by the presence of large scale
circulations induced by the temperature difference between day and night sides. One of the result of such mixing
is to produce complex abundance distribution which is homogenized longitudinally and thus quenched to the day side values,
while still having large variation over evening and morning limbs for some molecules like, e.g., \chhhh\ \citep{2012A&A...548A..73A,2014A&A...564A..73A},
thus demonstrating that 3D effects can have strong global impact on the atmospheric chemistry \citep{2020arXiv200103668M,2018ApJ...855L..31D,2018ApJ...869...28D}.
In emission observations, the effect of atmospheric winds may not be particularly important due to a rather large flux contrast
between day and night sides \citep[e.g.,][]{2018ApJ...869..107M}, 
however the thermal inversions in the upper atmospheric layers may result in noticeable
differences in day and evening sides of the planet that could possibly alter its disk integrated spectrum \citep{2014A&A...564A..73A}.
In transmission, to the contrary, the effect of circulations may play an important role in rendering
the observed spectra because it probes the terminator region where the differences in atmospheric 
thermal structure and physico-chemical processes determined by T(p)
are expected to be maximal and that, if ignored, could introduce biases in the retrieved
atmospheric parameters \citep{2019A&A...623A.161C}.}

\subsection{Planetary parameters}
We considered only one case of a Jupiter size planet orbiting stars of different temperatures. Our intention was to study a general trend of changes
in the predicted spectra caused by disequilibrium chemistry rather than to investigate particular details of the mixing ratio profiles as a function
of planetary parameters. For instance, planets more massive than one Jupiter mass would have denser atmospheres 
and hence less efficient turbulent mixing. Also, the penetration depth of the XUV radiation would be smaller.
An extensive investigation of these effects was recently carried out by \citet{2019ApJ...873...32M} 
and we refer the interested reader to this work for more details.

\section{Summary}

In this work we investigated how  disequilibrium chemistry impacts
the observed spectra and molecular mixing ratios in the atmosphere of a hot Jupiter (HJ)
orbiting stars of different types.
We also explicitly addressed the impact of stellar activity on the derived mixing ratios
and spectra simulated at primary and secondary eclipses. 
We demonstrated how these processes cause different trends in the predicted spectra, 
and identify major factors that can effect interpretation of observations.
The main conclusions of this work are summarized below.

\begin{itemize}
\item
Photodissociation by external stellar radiation and chemical kinetic drive mixing ratios
far from their thermo-equilibrium values. This effect is stronger for planets orbiting hot
stars because those generate stronger XUV fluxes compared to stars of later spectral types.
For planets having equilibrium temperatures around $\teq=1000$~K the effect of disequilibrium chemistry 
for some molecules (\hcn, \nhhh, \chhhh) can be seen even in optically thick layers around 1~bar and deeper,
in agreement with previous studies \citep[e.g.,][]{2014RSPTA.37230073M}.
\item
In all cases considered the \co\ remains the dominant carbon bearing molecule in the atmosphere.
However, for planets with $\teq=1000$~K that orbit G2 and K0 stars the concentration of \chhhh\ becomes comparable 
to that of \co\ at P$<$0.1~mbar, whereas 
at higher altitudes, a smaller XUV radiation from the K0 star 
allows for rather similar \co\ and \chhhh\ profiles throughout the atmosphere
given the inefficient methane photodissociation.
\item
We find that the effect of disequilibrium chemistry on the emission an transmission spectra 
is better observed for planets with  $\teq=1000$~K for all stellar spectral types,
thus favoring planets with relatively cool temperatures for the in-depth analysis
of disequilibrium effects \citep{2016SSRv..205..285M}. We find, in addition, that disequilibrium chemistry
has strong impact on the emission and transmission spectra of ultra hot planets 
with temperature inversion at high altitudes and that orbit A-type stars only.
For planets with  $\teq=2000$~K the observational consequences of disequilibrium chemistry on the spectra are not evident.
\item
In transmission, the most prominent features due to disequilibrium chemistry are the \nhhh\ bands at 1~\mum,
1.3~\mum, 1.5~\mum, and 2~\mum, as well as the strengthening of the \co/\coo\ absorption at 4.3~\mum\ and \nhhh\ absorption at 10~\mum.
We also notice a weakening of the \chhhh\ band at 3.3~\mum. Finally, in our calculations the disequilibrium chemistry leads to the appearance 
of the narrow \hcn\ absorption at 14~\mum\ which never shows up in equilibrium models. 
{The appearance of this \hcn\ feature has been noted in some previous works \citep[e.g.,][]{2011ApJ...737...15M} 
and its observation could potentially provide benchmark test for modern disequilibrium models.}
\item
The enhanced XUV radiation plays a crucial role in shaping mixing ratios of most atmospheric
species around stars with chromospheric activity, having strongest impact
in planets orbiting stars of young ages that have highest level of XUV radiation. 
However, even in the case of the present Sun with its known and
relatively low activity level, accounting for the true non-thermal XUV flux in disequilibrium calculations
is needed to predict realistic concentrations of species at high atmospheric layers.
On the other hand, in HJs orbiting stars similar to our Sun (in term of temperature and age),
the stellar XUV flux does not penetrate in line forming regions of planetary atmospheres and thus does not
affect their spectral appearance much.
\item
Spectra from future facilities with spectroscopic capabilities in
infrared promise to identify disequilibrium chemical constituents and their underlying kinetic controlling mechanisms.
Using the NIRSpec and MIRI instrument onboard of the JWST it will be possible to detect and analyze photochemically induced
changes in the observed spectra of hot exoplanets, {in agreement with estimates by \citet{2018ApJ...853..138B}}. 
In particular, the typical changes that we predict in this study
are observed with sufficiently low noise level in only one transit around a sun-like star with a K-band magnitude of 5$^m$.
{Future ESA mission ARIEL will also have spectroscopic capabilities. Although the detection
of disequilibrium chemistry could in principle be possible within 3$\sigma$ confidence interval
in HJs around G-type stars, the robust analysis would likely be not possible
mainly due to the low spectroscopic resolution of the instrument.}
Planet-hunting missions such as TESS\footnote{\tt https://heasarc.gsfc.nasa.gov/docs/tess/}, 
PLATO\footnote{\tt http://sci.esa.int/plato/}, and CHEOPS\footnote{\tt http://sci.esa.int/cheops/} 
are expected to significantly increase the number of exoplanets that orbit bright stars that will be accessible
for detailed atmospheric characterizations.
\item
Similar to many previous works, we ignored the effect of disequilibrium chemistry on the temperature
structure of planetary atmospheres. This effect is expected to be important at least in some temperature
regimes \citep{2016A&A...594A..69D} and should be taken into account in future works.
\end{itemize}

\begin{acknowledgement}
Authors thank the anonymous referee for insightful comments and suggestions to the manuscript.
Authors also wish to thank Matej Malik for his help with \helios\ and \heliosk\ codes and Natasha Batalha for her help with PandExo simulation package.
The estimates of exposure times for the ARIEL mission were kindly provided by Lorenzo Mungai and Enzo Pascale.
We additionally thank Bandow Bernhard for the help installing CUDA and Paul Hartogh for helpful discussions.
We acknowledge the support of the DFG priority program SPP-1992 “Exploring the Diversity of Extrasolar Planets” (DFG PR 36 24602/41).
LML acknowledges the financial support from the State Agency for Research of the Spanish MCIU
through the ``Center of Excellence Severo Ochoa'' award SEV-2017-0709, and
from the research projects ESP2016-76076-R and PGC2018-099425-B-I00.
We also acknowledge the use of electronic databases SIMBAD and NASA's ADS.
\end{acknowledgement}

\bibliographystyle{aa}
\bibliography{biblio}

\begin{appendix}

\end{appendix}

\listofobjects

\end{document}